\documentclass[]{elsarticle}

\pdfoutput=1

\usepackage[utf8]{inputenc}

\usepackage{amssymb, amsmath, amsthm}
\theoremstyle{plain}

\usepackage{siunitx}
\usepackage{graphicx}

\usepackage{url}
\usepackage[british]{babel}
\usepackage[nodayofweek]{datetime}

\usepackage{xcolor}
\usepackage[colorlinks]{hyperref}

\journal{Advances in Water Reseources}

\makeatletter
\def\ps@pprintTitle{%
 \let\@oddhead\@empty
 \let\@evenhead\@empty
 \def\@oddfoot{\hfill\footnotesize{\textit{\(\mathit 3^{\mathit{rd}}\) February, 2016}}}%
 \let\@evenfoot\@oddfoot}
\makeatother

\begin{document}

\begin{frontmatter}

\title{Building a Bridge from Moments to PDF's: A New Approach to Finding PDF Mixing Models}

\author[a,b]{L. Schüler\corref{1}} \ead{lennart.schueler@ufz.de}
\cortext[1]{Corresponding author.}
\author[c,d]{N. Suciu}
\author[c]{P. Knabner}
\author[a,b]{S. Attinger}
\address[a]{Institute of Geosciences, Friedrich Schiller University Jena\\ Burgweg 11, 07749 Jena, Germany}
\address[b]{Department Computational Hydrosystems, Helmholtz Centre for Environmental Research - UFZ,\\ Permoserstra{\ss}e 15, 04318 Leipzig, Germany}
\address[c]{Mathematics Department, Friedrich-Alexander University of Erlangen-Nuremberg,\\ Cauerstra{\ss}e 11, 91058 Erlangen, Germany}
\address[d]{Tiberiu Popoviciu Institute of Numerical Analysis, Romanian Academy,\\ Fantanele 57, 400320 Cluj-Napoca, Romania}

\begin{abstract}
Probability density function (PDF) methods are a promising alternative to predicting the transport of solutes in groundwater under uncertainty. They make it possible to derive the evolution equations of the mean concentration and the concentration variance, used in moment methods. A mixing model, also known as a dissipation model, is essential for both methods. Finding a satisfactory mixing model is still an open question and due to the rather elaborate PDF methods, a difficult undertaking. Both the PDF equation and the concentration variance equation depend on the same mixing model. This connection is used to find and test an improved mixing model for the much easier to handle concentration variance. Subsequently, this mixing model is transferred to the PDF equation and tested. The newly proposed mixing model yields significantly improved results for both variance modelling and PDF modelling.
\end{abstract}


\end{frontmatter}

\newpage
\section{Introduction}

%broad intro
Predicting the transport of groundwater contaminations remains a demanding task, especially with respect to the heterogeneity of the subsurface, the large measurement uncertainties, and the increasing impact of human activities on groundwater systems \cite{WWAP_united_2012}. Hence, a risk analysis also includes the quantification of the uncertainty in order to evaluate how accurate the predictions are.

Transport of dissolved contaminants in the subsurface is strongly influenced by the flow velocity of the groundwater, which in turn is determined by the properties of the surrounding geological structures. Their properties like the hydraulic conductivity are generally highly heterogeneous on many different scales.
%should this be mentioned?
These heterogeneities range from the order of magnitude of individual grains to large geological structures like facies, fractures and sediment layers.

%heterogeneities
Spatial fluctuations of the aquifer properties make the transport of the contaminants highly heterogeneous too. Water parcels transporting a contaminant and travelling very closely together can be separated and follow different and distinct flow paths. As a consequence, an enhanced spreading of the plume is observed. The impact of the heterogeneities on the transport behaviour is well known in general \cite{gelhar_three_1983, burr_nonreactive_1994}. But in order to predict the transport of contaminants accurately, it is necessary to know the heterogeneous aquifer properties influencing the transport everywhere. Monitoring the complete aquifer on all relevant scales is not feasible, but it is possible to retrieve partial knowledge by local measurements.
%heterogeneities -> stochastic description
These measurements can be used to generate a geostatistical representation of the aquifer. This way, the remaining uncertainty of aquifer properties and model parameters can be taken into account. For applying a stochastic framework, an ensemble of aquifer realisations is generated in accordance with the geostatistical representation. Hydraulic properties are modelled as spatial random functions, which in turn leads to the contaminant concentrations being modelled as spatial random functions too.

%ensemble avged description

\begin{figure}
 \centering
 \includegraphics[width=0.8\linewidth]{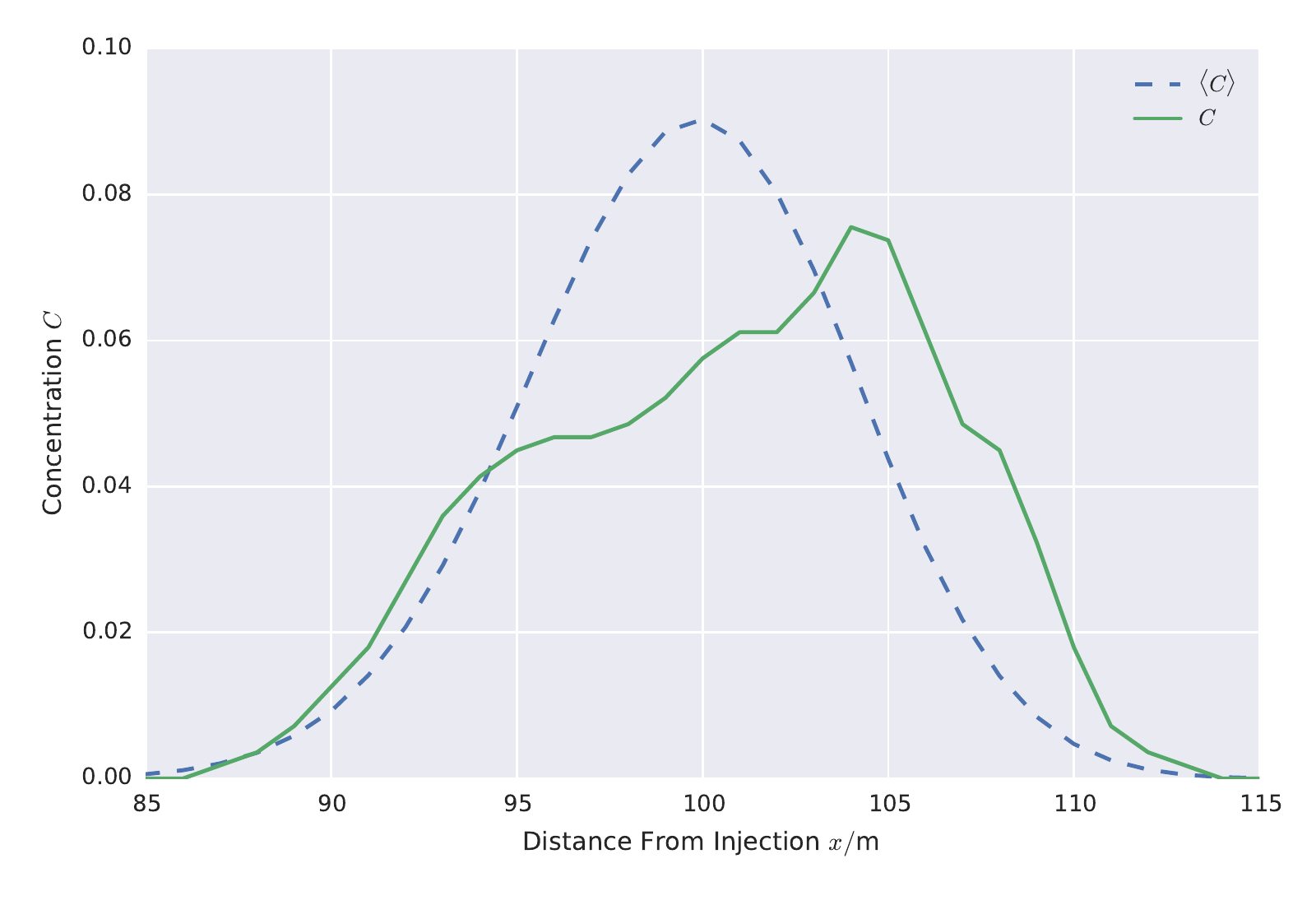}
 \caption{A measure is needed to quantify how good the mean concentration \(\langle C \rangle\) approximates the actual concentration \(C\), since the difference can be significant.}
 \label{fig:c-mean_c}
\end{figure}

The moment approach uses transport equations of the concentration moments consistent with the geostatistical representation of the aquifer's heterogeneity. If this heterogeneity is statistically homogeneous, the equation for the first moment, which is the mean concentration, has the following characteristics: The highly heterogeneous and spatially fluctuating groundwater velocity is replaced by an ensemble averaged velocity field and the effect of the fluctuating velocity on the transport is modelled by an enhanced dispersion called macrodispersion or ensemble dispersion \cite{gelhar_three_1983}.
%disadvantages of ensemble avged description
This approach has the limitation that the ensemble averaged concentration only describes the mean plume behaviour. In general, the mean behaviour differs from that of a specific plume in a single aquifer. See Figure \ref{fig:c-mean_c} for a comparison between the mean concentration and a concentration obtained from a simulation in a specific velocity field realisation. Only if such a single plume has sampled a representative part of the aquifer, it becomes ergodic and its transport behaviour can be modelled by the ensemble averaged behaviour, described above.
%variance as a measure
In a first step, possible deviations from the mean behaviour can be quantified by the concentration variance. It is transported by the same processes as the mean concentration, thus it is advected by the averaged velocity field and dispersed by an enhanced macrodispersion. But concentration variance is also generated by mean concentration gradients and simultaneously it is destroyed by dissipative processes, which are created by small-scale fluctuations in the velocity field. 
%closure model for variance
In order to calculate the influence of these small-scale fluctuations on the concentration variance, a so-called closure model is needed.

In the field of turbulence modelling, where very similar transport equations are used, different approaches exist for such closure models \cite[e.g.][]{tennekes_first_1972}. Up to this point, the adoption of these approaches to groundwater transport modelling has been hampered by the vastly different flow conditions prevalent in both fields. Contrary to most other problems where turbulent flows are more challenging, the roles are reversed here.
%difference turb. flows <-> porous media flows
The strong mixing induced by turbulent flows causes this closure problem to be easier to tackle. The mixing induced by heterogeneities in the groundwater flow is slower and changes significantly in time and is therefore more difficult to model.
%effective dispersion
As Dentz et al. \cite{dentz_temporal_2000-1} have shown, the mechanism which generates the physical mixing in a given aquifer realisation is more reliably described by the effective dispersion coefficients in comparison to the ensemble dispersion coefficients, which correspond to the turbulent diffusion coefficients.
The effective dispersion is small at early times and increases only slowly with time. Therefore, concentration gradients at early times are steep and may remain steep for prolonged times, which in turn prevents the smoothing of concentration fluctuations and preserves concentration uncertainty. Andricevic \cite{andricevic_effects_1998} proposed a mixing mechanism based on a time variable effective length scale which, in principle, could be determined experimentally. Kapoor and Gelhar \cite{kapoor_transport_1994, kapoor_transport_1994-1} derived a transport equation for the concentration variance, including local dispersivity and macrodispersive transport. By neglecting the local dispersivity, the results from Dagan \cite{dagan_stochastic_1982} could be derived. But it was concluded that even very small local dispersivities create a qualitatively different behaviour compared to the zero local dispersivity case, as the local dispersivity is the only mechanism which can reduce the variance. They used an approach developed for turbulent flows to model the variance dissipation, created by the local dispersivity. Furthermore, analytical solutions for the long-time behaviour of the concentration variance were derived. These results were confirmed for globally integrated variances by numerical simulations \cite{kapoor_advection-diffusion_1997}.

%PDF motivation via risk analysis
If the predictions made by a contaminant transport model are to be used for risk analysis, even more information than the mean concentration and the variance is needed. Risk thresholds, regulated e.g. by an environmental agency \cite{WWAP_united_2012}, can only be factored in by the so-called exceedance probability \cite[e.g.][]{andricevic_evaluation_1996}. It depends on the complete one-point probability density function (PDF) of the concentration. The concentration variance, as discussed above, can only be used to calculate a first estimation of the exceedance probability \cite{de_barros_simple_2011}. Even if such an estimation would be an acceptable approximation, rare events or extreme values cannot be mapped by the mean concentration and the concentration variance alone. This limitation stems from the fact that by using only the first two statistical moments, namely the mean and the variance, a Gaussian shape for the concentration PDF is implied. Such a distribution is short-tailed and therefore excludes the possibility of rare events. The first studies applying such a PDF framework to the transport in groundwater used a beta distribution, fully characterised by two parameters, to fit the concentration PDF in a non-Gaussian way \cite{fiorotto_solute_2002}. But Srzic et al. \cite{srzic_impact_2013} concluded that beta-shaped PDF's only match the true PDF for low heterogeneities. As a consequence, a second approach - the PDF approach - is investigated in this study. This approach yields an equation for the whole PDF of the concentration and thus makes no assumptions about the shape of it. The crux of these PDF methods is finding a mixing model. We refer to Celis and Figueira da Silva \cite{celis_lagrangian_2015} for a recent review of mixing models. Meyer et al. \cite{meyer_joint_2010} simulated the concentration PDF for advection-dispersion processes tailored to the transport in groundwater without assuming a specific shape. The PDF transport equation was solved by particle simulations developed for turbulent flows. A need for better mixing models was identified \cite{pope_pdf_1985, fox_computational_2003, meyer_joint_2010}. In our recent publications, we derived consistency conditions in order to link PDF equations to Fokker-Planck equations for which efficient numerical schemes exist \cite{suciu_fokker-planck_2015, suciu_consistency_2015}. We derived mixing models from simulated concentration time series. These mixing models performed well, but a trajectory needs to be prescribed on which the concentration is sampled. Prescribing such a trajectory is only possible in special cases. In addition, we used spatially filtered probability density functions (FDF) to further reduce the computational costs \cite{suciu_towards_????}.

%reaction term
Another major advantage of the PDF approach is the possibility to include mass transfer, like chemical reactions or radioactive decay, even in case of non-linear reactions. This intriguing property of the PDF approach is possible by assuming that the mass transfer solely depends on the concentration \cite{pope_pdf_1985, fox_computational_2003, suciu_consistency_2015}.

In this paper, we investigate the spatially resolved concentration variance and concentration PDF behaviour over a long time period and show that mixing models used before fail to reproduce the variance at all times. To that end, we first introduce a closed transport equation for the one-point concentration PDF in section \ref{sec:methods}. Using this equation as a starting point, we derive the transport equations of the first two statistical moments. We show that by prescribing a certain mixing model, both the PDF and the variance equations have the same closure problem and thus depend on the same mixing coefficients. The importance of this finding lies in the possibility of testing new mixing models with much simpler concentration variance simulations first and subsequently transferring them to PDF models. Next, we present analytical solutions of the moment equations and show the dependence of the analytical solution of the concentration variance on the mixing model. In section \ref{sec:IEM} we identify the need for a time dependency of the coefficients of the mixing model and we propose a new time dependent mixing model. This new model is explicit and in a closed form. It is then verified in section \ref{sec:simulations} by comparing the previously derived analytical solution of the concentration variance equation with the old and the new mixing model with reference Monte Carlo simulations. Afterwards, the new model is also used in the PDF framework and compared to reference Monte Carlo solutions. Finally, we conclude our work in section \ref{sec:conclusions}.

\section{Background}
\label{sec:methods}

The non-reactive transport of a solute in groundwater can be described by

\begin{equation}\label{eq:C}
 \frac{\partial}{\partial t}C + V_i \frac{\partial}{\partial x_i} C = D_{ij} \frac{\partial^2}{\partial x_i \partial x_j} C \; ,
\end{equation}
where \(C(\mathbf x, t)\) is the concentration of the solute which is transported by the statistically homogeneous random velocity field \(\mathbf V(\mathbf x)\) and the local dispersion \(\mathbf D\). It is assumed to be diagonal with \(D_{11} = D_L\) being the longitudinal component, \(D_{ii} = D_T\) for \(i>1\) being the transversal components, and \(D_{ij} = 0\) for \(i \neq j\). Throughout this work, we will be using the Einstein summation convention. The stochastic partial differential equation \eqref{eq:C} describes the time evolution of a plume of a dissolved substance in the groundwater. If an ensemble of statistically equivalent solutions of this equation is calculated, the mean behaviour can be calculated from the ensemble average \(\langle C\rangle = \sum_{i=1}^N C_i/N\) over \(N\) realisations. As a first measure, the variance \(\sigma_c^2(\mathbf x, t)\) can quantify how good the ensemble average approximates the behaviour in a specific realisation. The mean concentration \(\langle C \rangle(\mathbf x, t)\) and the concentration variance \(\sigma_c^2(\mathbf x, t)\) are the first and second statistical moments of the one-point one-time concentration PDF \(P(c; \mathbf x, t)\):

\begin{align}
 \langle C \rangle &:= \int c P \mathrm d c \label{eq:mean_c_def}\\
 \sigma_c^2 &:= \int c^2 P \mathrm d c - \langle C \rangle^2 \; .\label{eq:var_def}
\end{align}
Thus, if a transport equation for the PDF is derived, transport equations of the mean and variance can be derived too.

\subsection{PDF Transport Equation}
\label{sec:pdf_transport}
We have already shown the derivation of the PDF transport equation in two different ways in detail \cite{suciu_fokker-planck_2015, suciu_consistency_2015}. Hence, we only present the results here.

The PDF transport equation with a gradient-diffusion model applied can be formulated as

\begin{equation}\label{eq:PDF_transport}
 \frac{\partial}{\partial t} P + \left< V_i \right> \frac{\partial}{\partial x_i} P - D_{ij}^{\mathrm{ens}} \frac{\partial^2}{\partial x_i \partial x_j} P = - \frac{\partial^2}{\partial c^2} (\mathcal M P) \; ,
\end{equation}
where \(\left< \mathbf{V} \right>\) is the ensemble averaged velocity and \(\mathcal M (c, \mathbf x, t)\) the conditional dissipation rate, which is still unclosed. The ensemble dispersion coefficient tensor \(\mathbf D^{\mathrm{ens}}\) is diagonal with \(D_{11}^{\mathrm{ens}} = D_L^{\mathrm{ens}}\), \(D_{ii}^{\mathrm{ens}} = D_T^{\mathrm{ens}}\) for \(i>1\), and \(D_{ij}^{\mathrm{ens}} = 0\) for \(i \neq j\). In general, the ensemble dispersion coefficients \(\mathbf D^{\mathrm{ens}}\) are time dependent. But in the turbulent regime, these coefficients can be assumed to be constant, because mixing is so fast. In aquifers, the asymptotic values can be reached within a few advective time scales \cite{dentz_temporal_2000-1}.

The interaction by exchange with the mean (IEM) model for closing the mixing term was first formulated by Villermaux and Divillon \cite{villermaux_representation_1972} and by Dopazo and O'Brien \cite{dopazo_approach_1974} and still remains very popular for modelling reactive and turbulent flows \cite[see e.g.][]{colucci_filtered_1998, raman_consistent_2007, popov_implicit_2014}. It closes the mixing term by approximating it with

\begin{equation}\label{eq:M}
 M = \frac{\partial^2}{\partial c^2} (\mathcal M P) = -\frac{\partial}{\partial c} \left[ \chi \left( c - \langle C \rangle \right) P \right] \; ,
\end{equation}
where \(\chi\) is a parameter called the mixing frequency when used in PDF methods or the variance decay coefficient \cite{kapoor_transport_1994} when used in moment methods. It has to be prescribed and it will be discussed in detail in section \ref{sec:IEM}. This model causes concentration fluctuations to relax towards the local mean concentration in an exponentially decaying way.

\subsection{Mean and Variance Transport Equations}

The evolution equations of the mean concentration and the concentration variance can be derived from the PDF transport equation \eqref{eq:PDF_transport}. A detailed derivation is given in \ref{app:mean-var-deriv}, but the main ideas are also presented here.

The mean concentration is defined by equation \eqref{eq:mean_c_def}. Therefore, the PDF transport equation \eqref{eq:PDF_transport} is multiplied by \(c\) and integrated over the entire concentration space. The integral over the mixing term vanishes just as we would expect from equation \eqref{eq:M}. Now, the definition of the mean concentration \eqref{eq:mean_c_def} can be inserted and the well known advection-dispersion equation for the mean concentration of passive solutes in statistically homogeneous velocity fields is derived:

\begin{equation}\label{eq:mean_c_eq}
 \frac{\partial}{\partial t}\langle C \rangle + \langle V_i \rangle \frac{\partial}{\partial x_i} \langle C \rangle - D_{ij}^{\mathrm{ens}} \frac{\partial^2}{\partial x_i \partial x_j} \langle C \rangle = 0 \; .
\end{equation}

The concentration variance is defined by \eqref{eq:var_def} and thus, in order to derive the transport equation, the PDF evolution equation \eqref{eq:PDF_transport} with the mixing model \eqref{eq:M} is multiplied by \(c^2\) and integrated over the entire concentration space. The squared mean concentration \(\langle C \rangle^2\) appears after some manipulations. This term can be replaced by multiplying equation \eqref{eq:mean_c_eq} with \(\langle C \rangle\), which gives an equation for \(\langle C \rangle^2\). Now, the definition of the concentration variance can be plugged in, which results in the transport equation for the variance:

\begin{equation}\label{eq:pdf-2-var}
 \frac{\partial}{\partial t} \sigma_c^2 + \langle V_i \rangle \frac{\partial}{\partial x_i} \sigma_c^2 - D_{ij}^{\mathrm{ens}} \frac{\partial^2}{\partial x_i \partial x_j} \sigma_c^2 = 2 D_{ij}^{\mathrm{ens}} \frac{\partial}{\partial x_i} \langle C \rangle \frac{\partial}{\partial x_j} \langle C \rangle - 2 \chi \sigma_c^2 \; .
\end{equation}
It can be seen that the concentration variance is transported by the mean velocity \(\langle \mathbf V \rangle\) and by the ensemble dispersion coefficients \(\mathbf D^{\mathrm{ens}}\) exactly like the mean concentration. But in contrast to the transport equation for the mean concentration \eqref{eq:mean_c_eq} it also has a source and a sink term on the right hand side. The source term creates variance at mean concentration gradients and couples the two equations \eqref{eq:mean_c_eq} and \eqref{eq:pdf-2-var} weakly, as the coupling is only in one direction.
For this study, the most interesting term of equation \eqref{eq:pdf-2-var} is the last one on the right hand side. This sink term destroys variance by small-scale fluctuations. It is not closed and the same variance decay coefficient \(\chi\) as in the mixing term of the transport equation for the full PDF \eqref{eq:PDF_transport} appears here. This link makes it possible to test different propositions of the variance decay coefficient as a closure assumption for the transport equation for the concentration variance. Subsequently, the new proposition can be transferred to the PDF equation. The big advantage of testing different closures for the variance is that this equation is easier to handle. On the one hand, the variance equation has an analytical solution expressed by a time integral (see section \ref{sec:ana-sol}) which can be readily evaluated by numerical quadratures. And on the other hand, PDF equations are high-dimensional, with independent variables in both the physical and the concentration space and they have to be solved numerically.

\subsection{Analytical Solutions of the Moment Equations}
\label{sec:ana-sol}

With an analytical solution of the concentration variance transport equation, new mixing models can easily be examined and compared to Monte Carlo reference simulations. In order to make the analytical solutions for the first two moments easier, we assume that the ensemble dispersion coefficients \(\mathbf D^{\mathrm{ens}}\) are constant. The asymptotic value is therefore used. This assumption was already justified in section \ref{sec:pdf_transport}.

An analytical solution of the transport equation for the mean concentration \eqref{eq:mean_c_eq} can be found, for example, by transforming it into the frequency domain, which makes it an ordinary differential equation. A multivariate Gaussian distribution with zero mean and a diagonal covariance matrix \(2D_{ii}^{\mathrm{ens}} t_0\) is prescribed as the initial condition. It can be interpreted as a function which evolved from a Dirac delta function for a time span \(t_0\) according to equation \eqref{eq:mean_c_eq} without the advection term. The solution is then given by

\begin{equation}\label{eq:mean-c-ana}
 \langle C \rangle (\mathbf x, t) = \prod_{i=1}^{d} \left( 4 \pi D_{ii}^{\mathrm{ens}} (t+t_0) \right)^{-1/2} \exp\left(-\frac{(x_i - \langle V_i \rangle t)^2}{4 D_{ii}^{\mathrm{ens}} (t + t_0)}\right) ,
\end{equation}
where \(d\) is the spatial dimension.

Deriving an analytical solution of the concentration variance evolution equation is more involved than deriving a solution of the mean concentration equation. The most important steps of the derivation are outlined here, but a more detailed derivation is given in \ref{app:var-ana-deriv}. This derivation is similar to the one presented by Kapoor and Gelhar \cite{kapoor_transport_1994-1}, but we believe that the derivation presented here is easier to comprehend, because it is a more straightforward derivation by standard methods. Furthermore, we leave us the option open to include a time dependency of the variance decay coefficient \(\chi(t)\). The variance evolution equation \eqref{eq:pdf-2-var} is an inhomogeneous linear partial differential equation and as such we formulate a fundamental solution (also known as Green's function). The general solution of equation \eqref{eq:pdf-2-var} can then be calculated by the convolution of Green's function with the inhomogeneity \(2 D_{ij}^{\mathrm{ens}} \frac{\partial}{\partial x_i} \langle C \rangle \frac{\partial}{\partial x_j} \langle C \rangle\). Because the convolution transforms into a simple multiplication, it is transformed into Fourier space. Eventually, the solution of the mean concentration \eqref{eq:mean-c-ana} is needed. This is where the time shift \(t_0\) becomes important, because without it, a singularity would appear in the limit \(t \to 0\), as the Gaussian solution would tend to a Dirac delta function in this short time limit. By applying this time shift, the solution stays Gaussian and the singularity vanishes. Finally, the convolution can be calculated and an analytical solution is the result:

\begin{align}
 \sigma_c^2 (\mathbf x, t) = &\sum_{i=1}^d 2 D_{ii}^{\mathrm{ens}} \int\limits_0^t \mathrm d t' \prod_{j=1}^d \frac{\exp\left( -\frac{(x_j - \langle V_j \rangle t)^2}{2 D_{jj}^{\mathrm{ens}} (2t + t_0 - t')} \right)} {\left[(2 \pi D_{jj}^{\mathrm{ens}})^2 (2t + t_0 - t')(t'+t_0) \right]^{1/2}} \notag\\
 &\left[ \frac{t-t'}{2 D_{ii}^{\mathrm{ens}} (2t + t_0 - t')(t'+t_0)} + \frac{(x_i - \langle V_i \rangle t)^2}{\left(2 D_{ii}^{\mathrm{ens}} (2t + t_0 - t')\right)^2} \right] \notag\\
 &\exp\left( -2 \int_{t'}^t \mathrm d t'' \chi(t'') \right) . \label{eq:var-ana}
\end{align}
The time integral is rather well behaved and can easily be solved, for example by adaptive numerical quadrature algorithms. Kapoor and Gelhar \cite{kapoor_transport_1994-1} have further tackled this integral with \(\chi = \mathrm{const}\) by applying some long term approximations and came up with a closed analytical solution. But because the short time behaviour is of interest in this work, we will stay with solution \eqref{eq:var-ana}. The variance decay coefficient \(\chi\) appears in the argument of the last exponential function. Hence, new mixing models can be verified with this equation if, for example, compared to Monte Carlo reference solutions.

\section{A Time Dependent Extension of the IEM Model}\label{sec:IEM}

The IEM model describes the decrease of the concentration PDF too slow, as we have already pointed out \cite{suciu_fokker-planck_2015, suciu_towards_????}. It was developed for simulating turbulent flows. One major difference between turbulent flows and flows in porous media is the time scale on which mixing takes place. In the turbulent regime, it is often taken as a constant. And even there, a mixing time scale as a variable parameter has already been taken into account \cite{sabelnikov_extended_2006, jones_les_2012, dodoulas_large_2013}.

The original IEM model for turbulent flows approximates the conditional dissipation rate by equation \eqref{eq:M}. The variance decay coefficient \(\chi\) is proportional to the inverse mixing time scale. In classical PDF approaches, the latter is usually assumed to be proportional to the turbulence time scale \cite{pope_pdf_1985, celis_lagrangian_2015}. In large eddy simulations (LES), the mixing time scale is often estimated as a velocity \cite{dodoulas_large_2013} or as a diffusion time scale \cite{colucci_filtered_1998}. Colucci et al.  \cite{colucci_filtered_1998}, for instance, used the subgrid length scale \(\Delta_l\), which defines the transition from resolved to unresolved scales and the subgrid diffusion coefficient, which corresponds to an isotropic ensemble dispersion coefficient \(D^{\mathrm{ens}}\) in groundwater flows. Following this approach, we can formulate the variance decay coefficient as

\begin{equation}\label{eq:IEM}
 \chi = k_{\chi} \frac{D^{\mathrm{ens}}}{\Delta_l^2} \; .
\end{equation}
The dimensionless model parameter \(k_{\chi}\) is usually in the range of \(1 \leq k_{\chi} \leq 3\) \cite{pope_pdf_1985, colucci_filtered_1998}.

For groundwater systems, characterised by the anisotropic local dispersion coefficients \(D_{ij}\), Kapoor and Gelhar \cite{kapoor_transport_1994} arrived at a very similar equation for the variance decay coefficient, by introducing the Taylor microscales \(\Delta_{c_i}\), which characterise the gradients of the concentration fluctuations along the ith coordinate. The resulting variance decay coefficient is

\begin{equation}\label{eq:KIEM}
 \chi = \sum_{i,j=1}^d\frac{D_{ij}}{\Delta_{c_i}\Delta_{c_j}} \; .
\end{equation}
But the Taylor microscales could only be fitted to measurements, as a closed formula was not given.

We should recall that the IEM model was developed to approximate the second derivative of the conditional dissipation rate with respect to \(c\) \eqref{eq:M}. The conditional dissipation rate is defined by

\begin{equation}\label{eq:cond-diss}
\mathcal M = \left< D_{ij} \frac{\partial C}{\partial x_i}\frac{\partial C}{\partial x_j} \middle| c \right>\; .
\end{equation}
Here, \(\langle A | B \rangle = \langle AB \rangle / \langle B \rangle\) denotes the conditional expectation of \(A\) given \(B\). The conditional dissipation rate \(\mathcal M\) depends on the squared concentration gradients, which clearly evolve in time. But the IEM model has no way of accounting for this evolution. It only takes the difference between the current concentration and the local mean concentration into account. As we have already observed \cite{suciu_towards_????}, a more accurate mixing model would account for larger dissipation rates at early times and smaller dissipation rates at later times, as the concentration gradients decrease. For turbulent reactive flows, a dependence of \(\chi\) on the Reynolds number of the subgrid scale flow was already proposed \cite{jones_les_2012, dodoulas_large_2013}. Furthermore, Sabel'nikov et al. \cite{sabelnikov_extended_2006} modelled the mixing frequency as a stochastic process in order to account for the entire range of time scales in the mixing process. They named their model extended interaction by exchange with the mean (EIEM). We elaborate the idea of using a variable variance decay coefficient \(\chi(t)\) and propose a new time dependent extension of the model, adapted to the transport processes in groundwater.
 
Like Andricevic \cite{andricevic_effects_1998}, we approximate the concentration gradients using an evolving effective spatial scale \(\lambda_c(t)\). This assumption implies that the squared concentration gradients evolve inversely proportional to that squared characteristic length scale \((\nabla C)^2 \sim \lambda_c(t)^{-2}\) as the plume spreads and its fringes and fluctuations smooth out. Since the concentration fluctuations are smoothed out by the local dispersion with the characteristic scale \(\lambda_c(t) = \sqrt{2Dt}\), we assume a decay of the conditional dissipation rate \eqref{eq:cond-diss} to be proportional to \( \mathcal M \sim \lambda_c(t)^{-2}\). In order to improve the IEM model, we include this dependency on \(\lambda_c(t)\) into the variance decay coefficient \eqref{eq:IEM}.

Furthermore, the ensemble dispersion coefficient \(\mathbf D^{\mathrm{ens}}\) accounts for an artificial dispersion which is caused by centre of mass fluctuations of the solute plume from realisation to realisation. The effective dispersion coefficient \(\mathbf D^{\mathrm{eff}}\) excludes this artificial dispersion and converges to \(\mathbf D^{\mathrm{ens}}\) in the long-time limit for velocity fields with short range correlations \cite{suciu_diffusion_2014, dentz_temporal_2000-1}. Because the mixing in turbulent flows is so much faster than it is in groundwater flows, the difference does not matter for turbulent flows. Therefore, the mathematically simpler to handle ensemble dispersion coefficient is used in studies concerning flows in the turbulent regime \cite{pope_pdf_1985, colucci_filtered_1998}. But in groundwater flows the difference is significant and because the centre of mass fluctuations do not influence the dissipation, the effective dispersion coefficient \(\mathbf D^{\mathrm{eff}}\) describes the correct behaviour for the mixing model. With these physical arguments and choosing \(k_{\chi} = 2\) from the middle of the interval of reported values, we propose following time-dependent variance decay coefficient:

\begin{equation}\label{eq:TD-IEM}
 \chi(t) = \sum_{i,j=1}^d \frac{D_{ij}^{\mathrm{eff}}(t)}{D_{ij} t}\; .
\end{equation}
In order to show the similarities between this newly proposed variance decay coefficient and the previous ones, we assume an isotropic correlation length of the log conductivity field \(\lambda\) and an isotropic local dispersion coefficient \(D\), which gives us the dispersive time scale \(\tau_D = \lambda^2 / D\). With this relationship, we can transform the variance decay coefficient to

\begin{equation}\label{eq:TD-IEM2}
 \chi(t) = \sum_{i,j=1}^d \frac{D_{ij}^{\mathrm{eff}}(t)\tau_{D}}{\lambda^2 t}\; .
\end{equation}
Equation \eqref{eq:TD-IEM2} generalises equation \eqref{eq:IEM} to a time-variable characteristic length scale and to anisotropic effective dispersion coefficients. Compared to the coefficient \eqref{eq:KIEM} introduced by Kapoor and Gelhar \cite{kapoor_transport_1994}, the new variance decay coefficient \eqref{eq:TD-IEM2} depends on the effective dispersion coefficients instead of the local dispersion coefficients. Furthermore, the unclosed Taylor microscale was replaced by the correlation length and a dimensionless time factor \(\tau_D / t\) was added.

\begin{figure}
 \centering
 \includegraphics[width=0.8\linewidth]{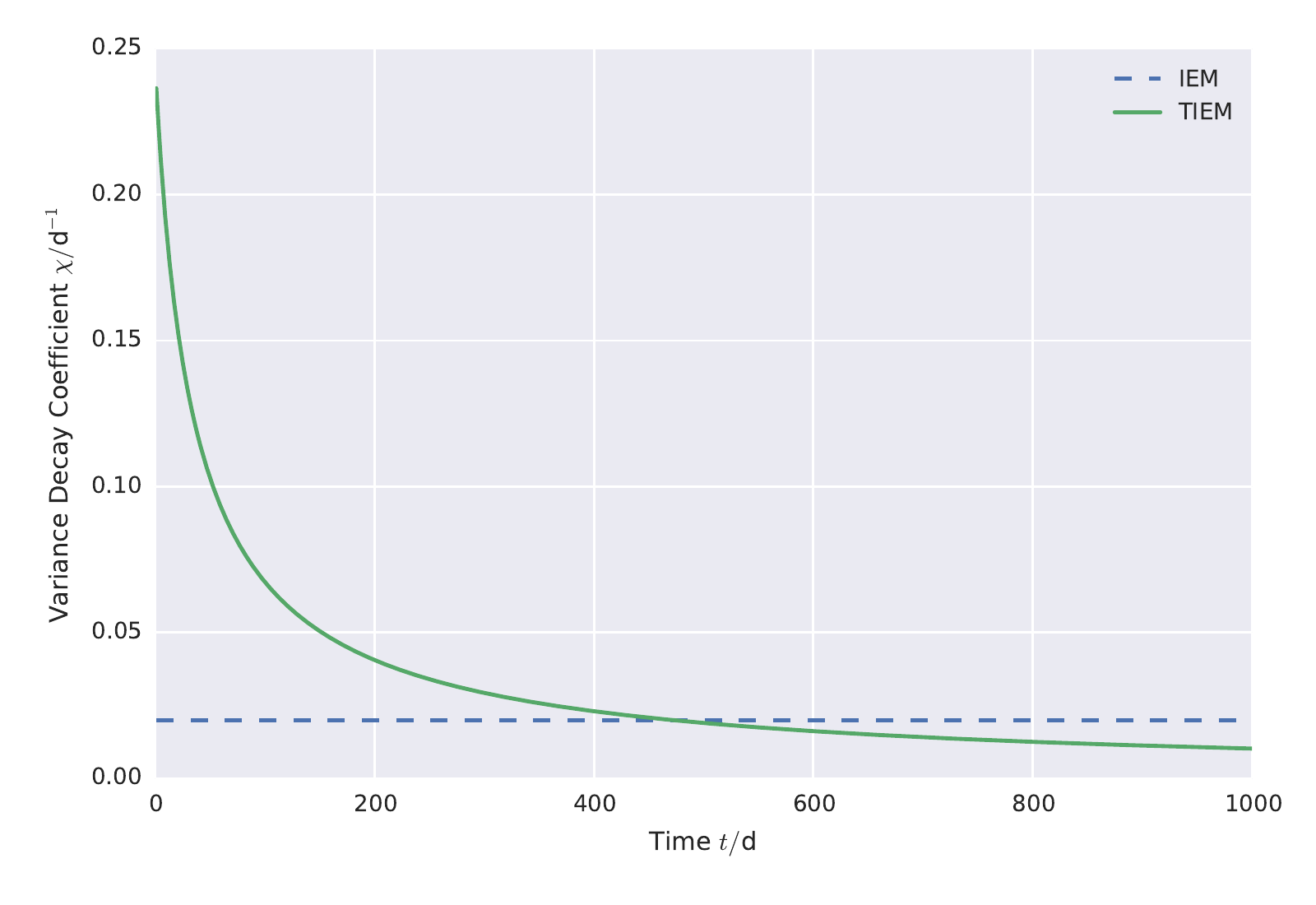}
 \caption{An illustration of the different time behaviours of the IEM and the TIEM variance decay coefficients. The TIEM model causes strong dissipation at early times, but for long times it causes less dissipation than the IEM model.}
 \label{fig:M}
\end{figure}

As shown in figure \ref{fig:M}, this variance decay coefficient has larger values than the constant one at early times, which causes a stronger dissipation. But then it drops below the constant variance decay coefficient and approaches \(\lim_{t \to \infty} \chi(t) = 0\). In order to distinguish this model from other extensions of the IEM model, we name it the time dependent interaction by exchange with the mean model (TIEM).

With this extension, the simplicity and low computational costs of the IEM model are preserved, while at the same time, it incorporates the time dependent physical processes causing the dissipation.

\section{Simulations}
\label{sec:simulations}

\subsection{Variance Modelling}
\label{subsec:var-modelling}

In order to verify the TIEM model independently, simulations with two different numerical models were performed. A sequential standard particle tracking model was implemented following Dentz et al. \cite{dentz_temporal_2002} and the global random walk (GRW) algorithm \cite{vamos_generalized_2003} was used as an independent model. The mean concentration and the concentration variance were derived from both numerical simulations and compared to the analytical solutions \eqref{eq:mean-c-ana} and \eqref{eq:var-ana} with the IEM and TIEM mixing models.

\subsubsection{Simulation Setup}
\label{subsec:num-sim}

Both numerical transport simulations were calculated for a two-dimensional heterogeneous velocity field which was modelled as a solution of the linearised Darcy and continuity equations by the Kraichnan algorithm \cite{kraichnan_diffusion_1970}. The mean flow velocity was prescribed as \(\langle \mathbf V \rangle = (1, 0)^T\SI{}{\metre\per\day}\) and an isotropic Gaussian covariance structure with a correlation length of \(\mathbf \lambda = (1, 1)^T\SI{}{\metre}\) and a variance of \(\sigma^2 = 0.1\) was chosen for the underlying conductivity field. The flow fields were generated by using 6400 Fourier modes for the randomisation method \cite{hese_generating_2014, suciu_towards_????}, in order to capture the self-averaging behaviour of the transport process over hundreds of correlation lengths \cite{eberhard_self-averaging_2007}.

The simulations were performed with an isotropic local dispersion coefficient of \(D = \SI{0.01}{\metre\squared\per\day}\). The particles were initially distributed uniformly on a rectangle with side lengths according to an initial dispersion time of \(t_0 = \SI{10}{\day}\) (see equation \eqref{eq:mean-c-ana}). Both numerical models used a time step of \(\Delta t = \SI{0.5}{\day}\).

For the standard particle tracking simulation, the particles, transported by the velocity field and performing random jumps, were modelled according to the It\^o equations

\begin{equation}\label{eq:langevin}
 \mathrm d X_i(t) = V_i(\mathbf X) \; \mathrm d t + \sqrt{2D} \; \mathrm d W_i(t) \; ,
\end{equation}
where \(W_i(t)\) are independent standard Wiener processes \cite{suciu_fokker-planck_2015}. An extended Runge-Kutta scheme \cite{drummond_scalar_1984} with an accuracy of order \((\Delta t)^{3/2}\) was used to discretise the stochastic equations \eqref{eq:langevin}.

1000 realisations with 150000 particles in each of them were calculated to create a statistical ensemble. It took about \(\SI{1100}{\minute}\) to compute one realisation on a single core of the EVE cluster at the UFZ Leipzig.

The GRW-algorithm takes a different approach. It uses a superposition of many weak solutions of It\^o equations projected onto a regular grid. The particles solving the It\^o equations are spread on the grid globally according to the drift and diffusion coefficients of the equation. By construction, this algorithm is free of numerical diffusion and can be used for practically arbitrary numbers of particles without an impact on the computational costs. For more details about the GRW algorithm, Suciu et al. \cite{suciu_numerical_2006} show how to implement an efficient GRW version of Monte Carlo simulations, whereas more technical details and the convergence behaviour of the schemes are presented by Suciu et al. and Suciu \cite{suciu_coupled_2013, suciu_diffusion_2014}. The same physical parameters were used as for the standard particle tracking. The GRW simulations where performed on a grid with \(4600 \times 1800\) cells with a resolution of \(\SI{0.1}{\metre} \times \SI{0.1}{\metre}\). A total of \(10^{24}\) particles were used to represent the behaviour of the concentration on the GRW lattice. The computation of the velocity field on the grid and the GRW transport simulation took about \(\SI{48}{\minute}\) for each realisation. The ensemble of realisations of the transport process was obtained by conducting independent simulations on 1000 cores, in a single job executed on the JURECA supercomputer at Research Centre Jülich.

A normalised two-dimensional histogram on grid cells with a size of \(\SI{1}{\metre} \times \SI{1}{\metre}\) was performed for both simulations to calculate concentrations from the particle distributions.

\begin{figure}
 \centering
 \includegraphics[width=\linewidth]{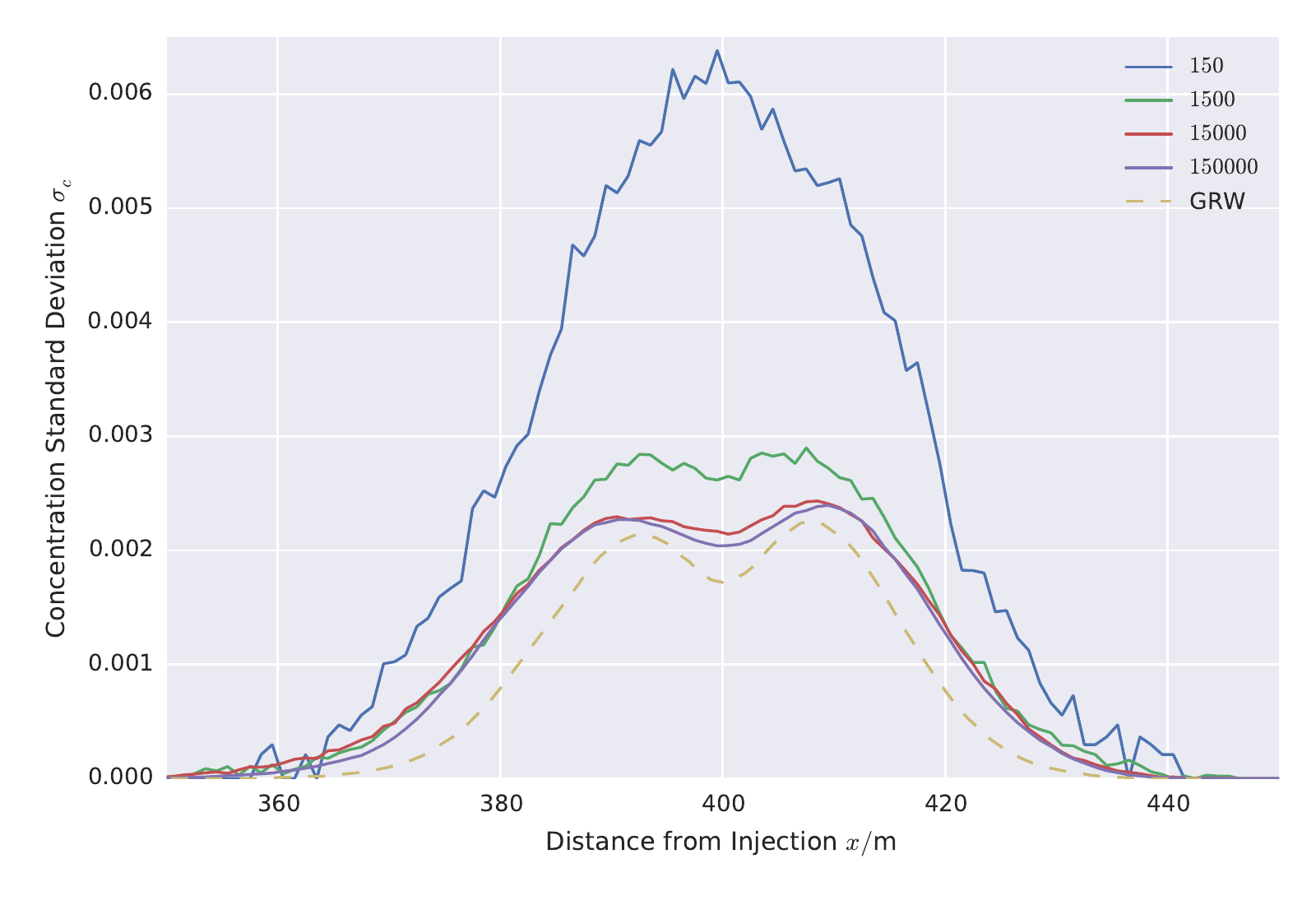}
 \caption{The concentration standard deviation calculated from standard particle tracking simulations with different amounts of particles per realisation compared to results from GRW simulations with \(10^{24}\) particles per realisation.}
 \label{fig:N}
\end{figure}

A comparison of the two numerical approaches shows how much the number of particles required for accurate simulations of localised quantities reduces the computational performance in classical, sequential particle tracking methods. Fewer particles are needed to compute global quantities, such as spatial moments of the solute plume \cite[e.g.][]{dentz_temporal_2002}. But for accurate estimations of the local variance of the solute concentration, \(10^5\), or even more particles are required (see Figure \ref{fig:N}). This results in a dramatic increase of computational time. Comparing the computing times normalised by the corresponding numbers of particles we find that GRW simulations are about \(10^{20}\) times more efficient in estimating the same localised quantity.

\subsubsection{Results}

Here, we investigate the impact of the TIEM model \eqref{eq:TD-IEM} on the analytical solution \eqref{eq:var-ana} of the concentration variance evolution equation \eqref{eq:pdf-2-var} by comparing the results to the two numerical models described in section \ref{subsec:num-sim}. Because no mixing term appears in the evolution equation for the mean concentration \eqref{eq:mean_c_eq}, different mixing models do not influence the mean concentration behaviour. Thus, the results for the mean concentration will not be shown here.

When using it in the analytical solution, the IEM model destroys variance globally, because it is space independent. Hence, the well-known bimodal shape of the variance of a Gaussian-like mean concentration will remain unaltered and only the magnitude of the variance will change by introducing new mixing models which act globally.

\begin{figure}
 \centering
 \includegraphics[width=\linewidth]{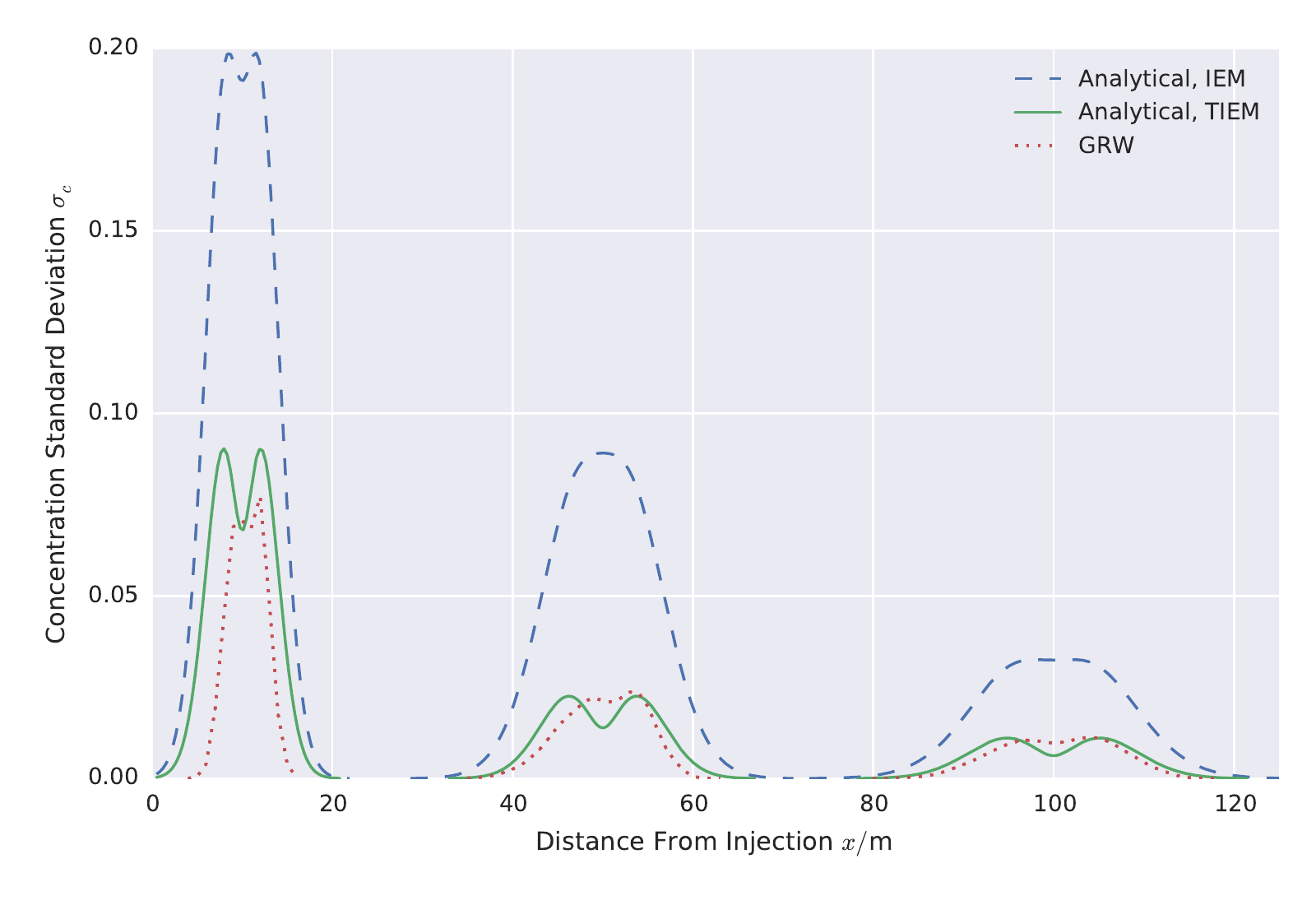}
 \caption{Analytical concentration standard deviations with the IEM and TIEM mixing models compared to concentration standard deviations computed from GRW simulations at times \(t=\) \SIlist{10;50;100}{\day}.}
 \label{fig:var_compare}
\end{figure}

\begin{figure}
 \centering
 \includegraphics[width=\linewidth]{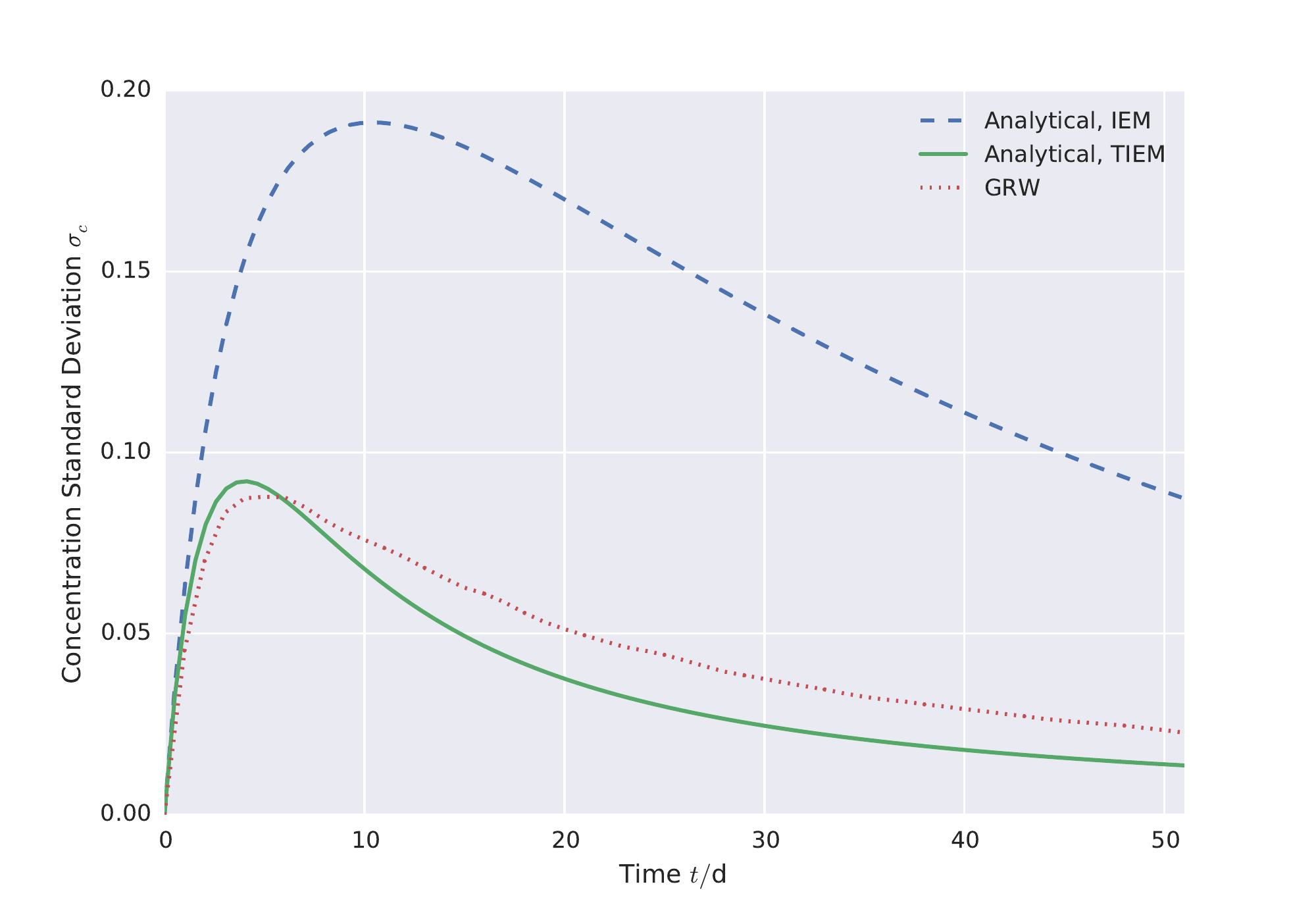}
 \caption{The concentration standard deviation at the centre of mass \(\mathbf x_{cm} = \langle \mathbf V \rangle t\).}
 \label{fig:std_cm}
\end{figure}

\begin{figure}
 \centering
 \includegraphics[width=\linewidth]{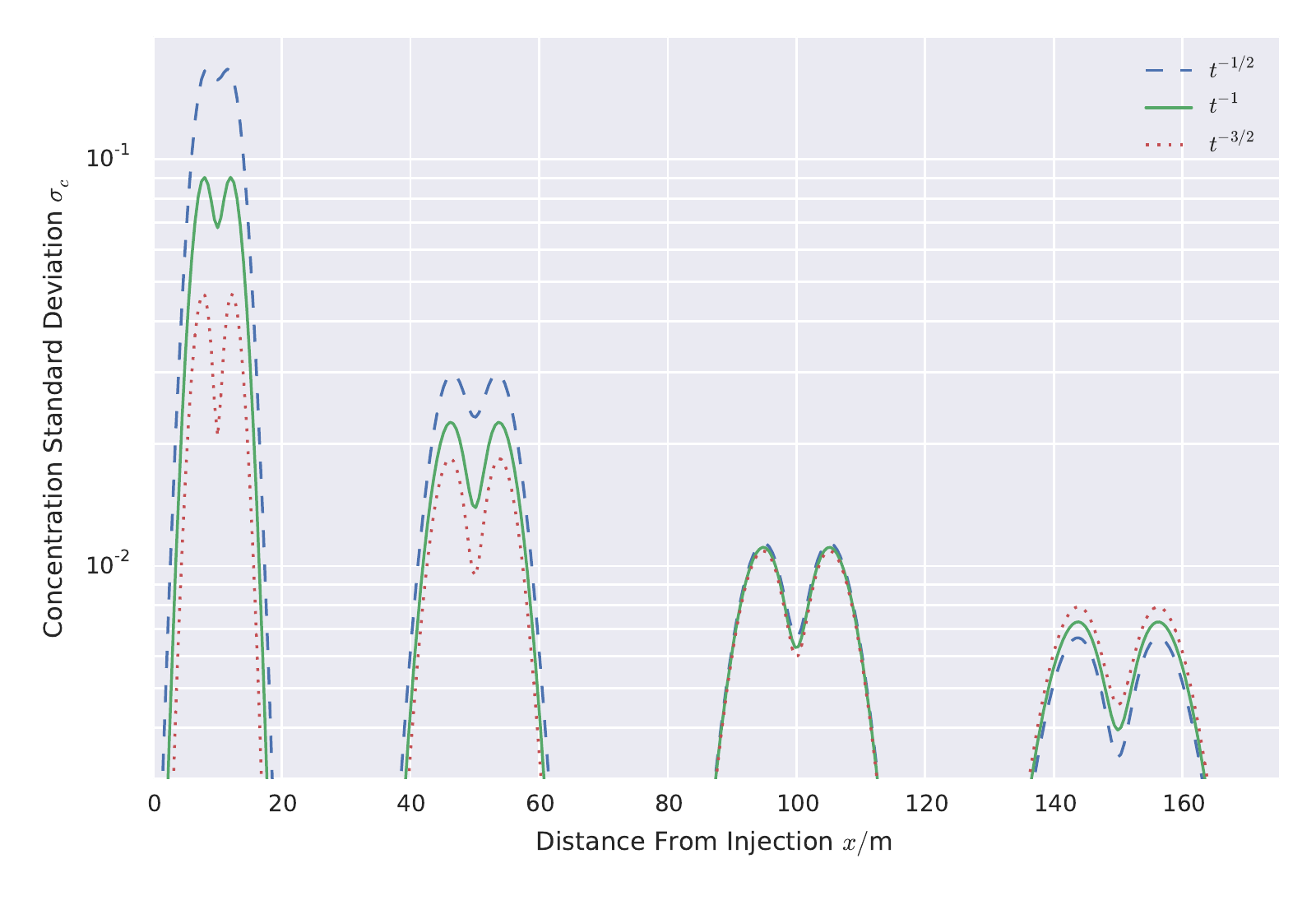}
 \caption{The analytical concentration standard deviation with different exponents \(a\) for \(t^a\). The solution with \(a=-1\) is the same as in figure \ref{fig:var_compare}. A semi-log plot is used in order to make the differences at \(t = \SI{150}{\day}\) clearer.}
 \label{fig:std_exponents}
\end{figure}

\begin{figure}
 \centering
 \includegraphics[width=\linewidth]{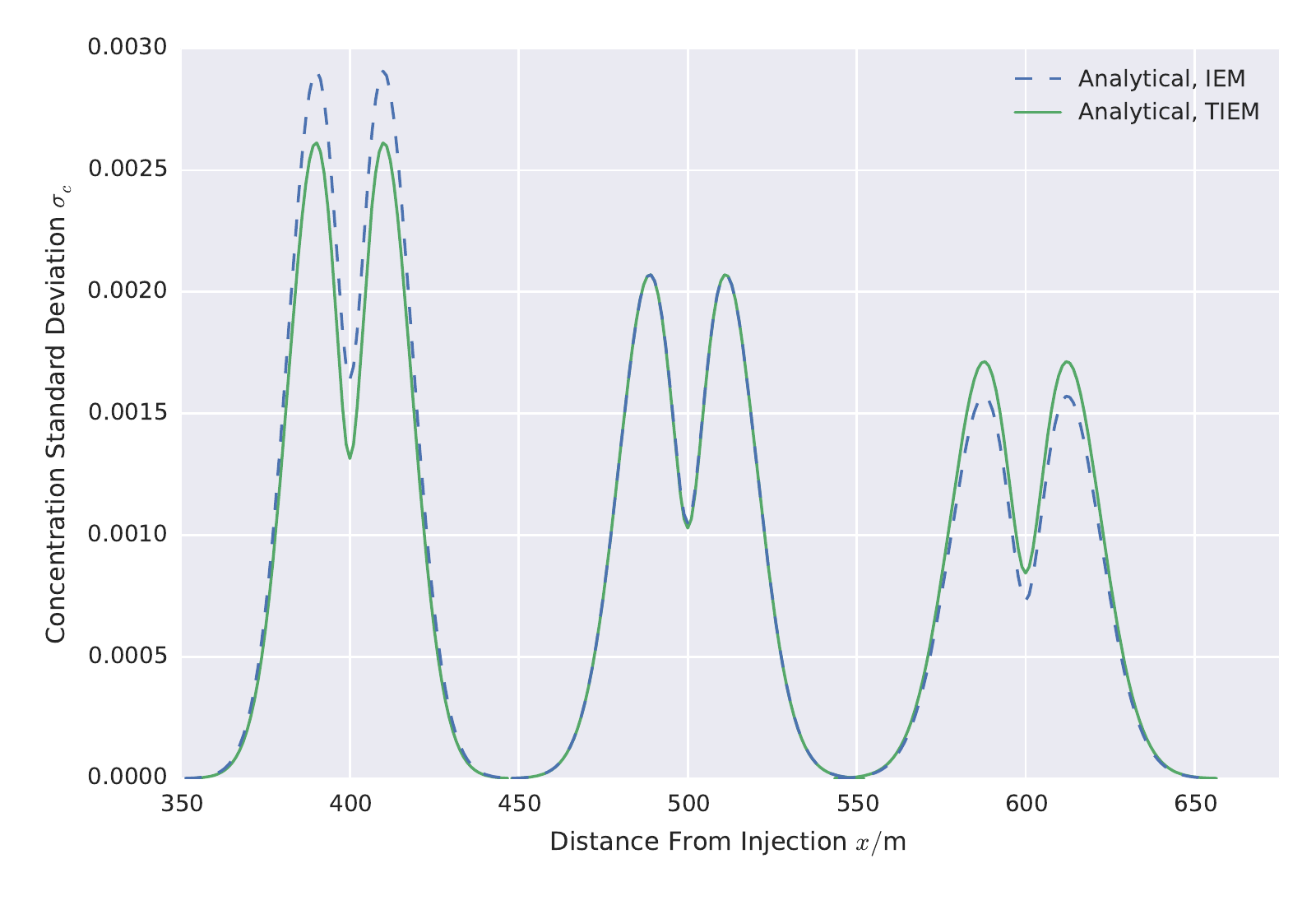}
 \caption{For \(t > \SI{500}{\day}\), the TIEM solutions stays larger than the IEM solution.}
 \label{fig:std-large-times}
\end{figure}
In Figure \ref{fig:var_compare}, the concentration standard deviation \(\sigma_c\) computed from the GRW simulation is compared to the analytical solution \eqref{eq:var-ana} using the two different mixing models. For the ensemble and effective dispersion coefficients, the results from Dentz et al. \cite{dentz_temporal_2000-1} where used. The results from the particle tracking are omitted, because they are very similar to the GRW solutions and make the figure difficult to read. The different solutions are plotted at \(t =\) \SIlist{10;50;100}{\day} after injection. Instead of the variance, its square root, the standard deviation, is plotted in Figure \ref{fig:var_compare}, because this quantity can be better compared to concentrations, than the variance, which is the squared deviation from the mean concentration. Furthermore, the already large differences of the curves at \(t=\SI{10}{\day}\) are squared when comparing the variances, making it difficult to compare them in a single graph.

The most obvious feature of the figure is the large peak of the analytical solution at short times with the IEM mixing model. This large peak shows the problem of the IEM model, namely that the variance destruction at short times due to small-scale fluctuations of the flow field is strongly underestimated. As seen from Figure \ref{fig:M}, the TIEM model has a much larger variance decay coefficient at short times which manifests itself as a stronger decline of variance at these early times. In Figure \ref{fig:var_compare}, this behaviour can be seen in the analytical solution with the TIEM model, which matches the numerical simulation very well at intermediate and long times. At \SI{500}{\day}, the analytical solutions with the IEM and the TIEM models intersect. At even larger times, the TIEM solution stays greater than the IEM solution, as shown in figure \ref{fig:std-large-times}. On the other side of the time axis, at very early times, ranging from \(t=\SI{0}{\day}\) up to about \(t=\SI{15}{\day}\) there is still a gap between the numerical reference simulations and the TIEM model. But for \(t = \SI{10}{\day}\) the IEM model differs from the reference simulations by about \(61\%\), compared to a difference of \(18\%\) with the TIEM model, which is a major improvement. The time evolution of the concentration standard deviation at the centre of mass of the mean concentration plume \(\mathbf x_{\mathrm{cm}} = \langle \mathbf V \rangle t\) is shown in Figure \ref{fig:std_cm}, to highlight the influence of the time dependent mixing model at early times. It should be noted that the analytical solution pronounces the valley of the bimodal structure of the variance curve too much. Comparing the peaks of the analytical solution and of the GRW solution, the difference at early times is more pronounced and becomes increasingly smaller at intermediate and long times. The slight asymmetry in the numerical solutions is due to the non-ergodicity at early times.

Finally, we tested the impact of different exponents of the explicit time dependency of the TIEM model \eqref{eq:TD-IEM2}. In Figure \ref{fig:std_exponents}, the standard deviation curves with the exponents \(t^{-1/2}\) and \(t^{-3/2}\) are compared to the exponent \(t^{-1}\), which follows from our arguments made in section \ref{sec:IEM}. It can be seen that the exponent of \(-1/2\) causes the variance to be too large at early times, which then decreases so fast over time, that it is less than the reference solution for \(t > \tau_D = \SI{100}{\day}\). Thus, even if the large values at early times would be adjusted by the constant factor \(k_{\chi}\) in the TIEM model, the variance would quickly drop beneath the reference values. On the other hand, the exponent of \(-3/2\) causes the variance to be too small at early times, which then decreases so slowly, that for \(t > \SI{100}{\day}\), it is greater than the reference solution. These results further support the physical reasoning made in chapter \ref{sec:IEM}.

\subsection{PDF Modelling}

As we have pointed out, Lagrangian particle methods used to solve PDF problems in the fields of combustion and turbulence are not well suited for groundwater problems, where concentrations are strongly diluted \cite{suciu_towards_????, suciu_consistency_2015}. Therefore, numerical simulations of the PDF equation with the new TIEM model where only performed with the GRW algorithm adapted to PDF simulations. The concentration PDF at the centre of mass of the plume was simulated based on the GRW setup described by us in \cite{suciu_fokker-planck_2015}. There, we showed that the PDF equation for the concentration at the centre of mass of the plume integrated over the transversal direction can be formulated as a two-dimensional Fokker-Planck equation. This equation describes the cross-section of the concentration at the centre of mass, for which corresponding Itô equations are formulated:

\begin{align}
 \mathrm d X(t) &= \langle V_1 \rangle \; \mathrm d t + \sqrt{2 D_{11}^{\mathrm{ens}}} \; \mathrm d W(t)\\
 \mathrm d C(t) &= M \; \mathrm d t \; ,\label{eq:pdf-ito-c}
\end{align}
These stochastic differential equations can be solved by Monte Carlo methods and thus by the GRW algorithm. The same parameters as for the simulations described in section \ref{subsec:var-modelling} were used. A reference solution was calculated from Monte Carlo simulations \cite{suciu_fokker-planck_2015}.

\begin{figure}
 \centering
 \includegraphics[width=\linewidth]{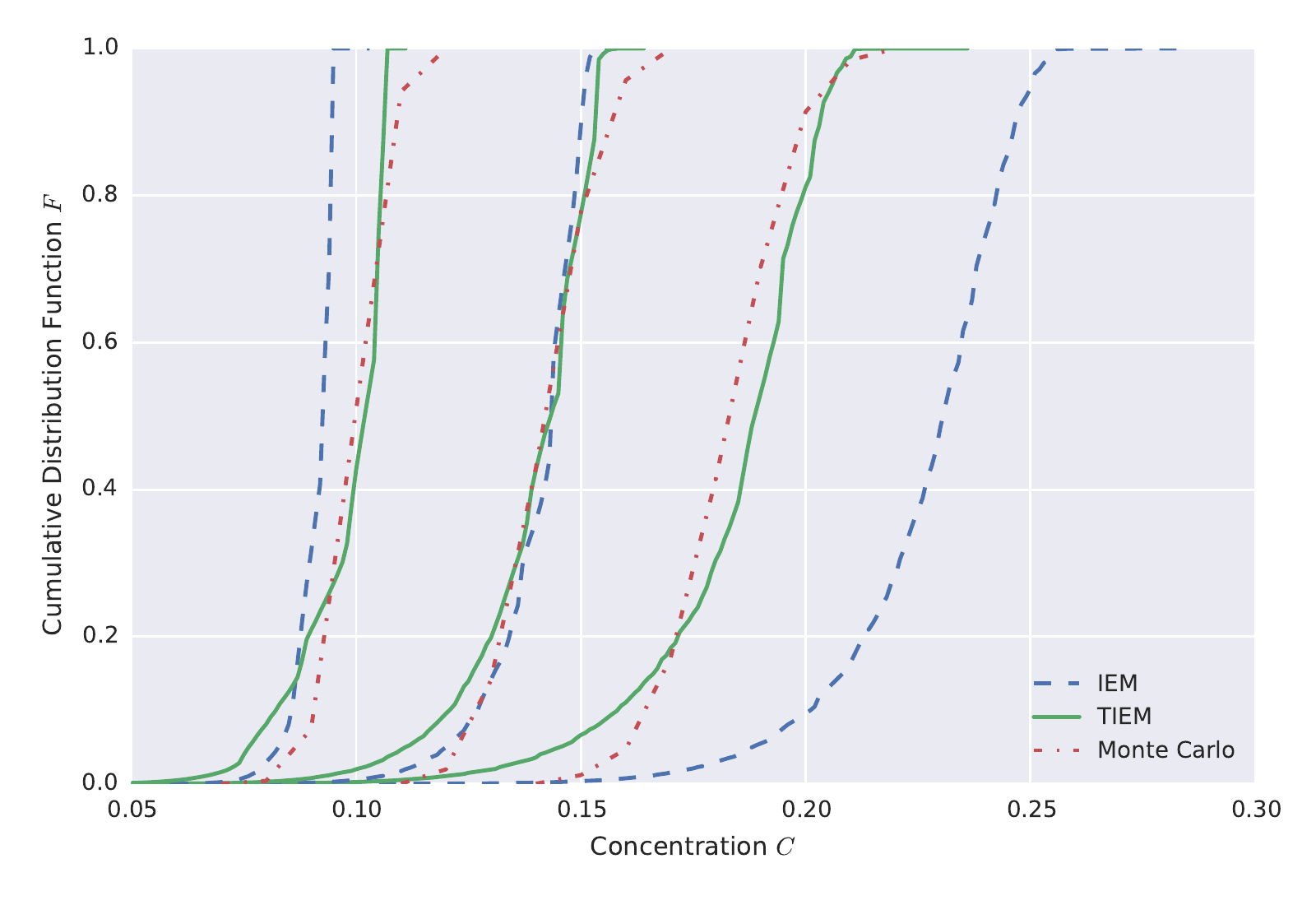}
 \caption{The CDF at the centre of mass of the solute plume at times \(t =\) \SIlist{30;50;100}{\day} (from right to left) is calculated with two different IEM models and with dissipation rates extracted from simulated particle trajectories.}
 \label{fig:cdf-mixing-mismatch}
\end{figure}

The results are shown in Figure \ref{fig:cdf-mixing-mismatch}. Here, the cumulative distribution function (CDF) \(F(c;x,t)\) and therefore the integral of the PDF is shown, because in general the CDF is a smoother curve than the PDF and can thus be better compared. The CDF at the centre of mass is shown at \(t =\) \SIlist{30;50;100}{\day} after injection (from right to left). It can be seen that the TIEM model is a major improvement over the IEM model. At early times, the IEM model predicts a CDF which is shifted far towards higher concentrations. The TIEM model is just slightly shifted, but the shape differs too with a longer tail towards low concentrations, similar to the IEM model. At \(t = \SI{50}{\day}\) both models perform acceptable. At \(t = \SI{100}{\day}\) the IEM model is even shifted too far towards lower concentrations, while the TIEM model is still close to the reference solution. The deviation of the IEM model from the reference solution indicates that the drift in concentration space (see equation \eqref{eq:pdf-ito-c}) is too slow at early times and too large at large times. By considering a time variable variance decay coefficient \(\chi(t)\) (see figure \ref{fig:M}), the TIEM model proposed in this paper provides a correction for the drift in concentration space. This leads to the observed improvements of the PDF simulations. Sabel'nikov et al. \cite{sabelnikov_extended_2006} also extended the IEM model to incorporate a time dependency of the variance decay coefficient for turbulent reactive flows. Compared to direct numerical simulations, they too reported a good match at intermediate times, but an increasing mismatch for small and large times.

\section{Conclusions}
\label{sec:conclusions}

This paper presents a new and time-dependent mixing model: an extended IEM model for groundwater, named TIEM. We showed that the same mixing model is used for both the concentration variance evolution equation \eqref{eq:pdf-2-var} and the concentration PDF evolution equation \eqref{eq:PDF_transport}. This link was used to verify the new TIEM model \eqref{eq:TD-IEM} with the much simpler to handle variance equation. The verification was done by comparing an analytical solution of the variance equation \eqref{eq:var-ana}, which depends on a mixing closure model, to two independent numerical models. The TIEM model shows a strong improvement over the IEM model. A significant deviation from the reference simulations can only be observed at times \(t < \SI{15}{\day}\). And even for these very short times, the new model is a significant improvement.

Based on these promising results, the model was transferred to the PDF framework. The results obtained from the PDF simulations with the TIEM model are not quite as satisfying as the results from the variance simulations mentioned above. Although there are mismatches at early and also at later times, the new model is still a major improvement over the classical IEM model. One possible way of further improving the IEM model is to derive a partial differential equation as a dynamic model for the variance decay coefficient \cite{im_large_1997, pierce_dynamic_1998}. Such a model could include the actual and instantaneous length scales of the processes destroying the variance. This feature would make it possible to also apply the model to statistically non-homogeneous conductivity fields \cite{pierce_dynamic_1998}, as needed if the fields are to be conditioned on measurements.

The GRW-simulations, together with the TIEM model, can easily be extended to three-dimensional problems, to anisotropic dispersion coefficients, and to reactive transport. Especially the latter point is worth highlighting. The reaction terms can simply be plugged into the PDF equations, which makes the PDF framework the method of choice for modelling complex reactive transport in groundwater.

\section*{Acknowledgements}
We kindly thank Veronika Zeiner, Malte Schüler, Jannis Weimar, and Falk Heße for their helpful suggestions and proof reading and David Schäfer for his help with parts of the numerical code. This work was supported by the Deutsche Forschungsgemeinschaft - Germany under Grant SU 415/2-1, KN 229/15-1, SU 415/2-1, by the J\"{u}lich Supercomputing Centre--Germany as part of the Project JICG41, and by the Bundesministerium für Wirtschaft und Energie as part of the H--DuR Project (funding number: 02 E 11062 B).

\appendix
\section{Mean and Variance Transport Equations}\label{app:mean-var-deriv}

The transport equation for the mean concentration can be derived from the PDF transport equation \eqref{eq:PDF_transport} by multiplying it with \(c\) and integrating over the entire concentration space. Doing so yields

\begin{equation}
 \int\limits_0^1 c \frac{\partial P}{\partial t} \;\mathrm d c + \int\limits_0^1 c \langle V_i \rangle \frac{\partial P}{\partial x_i} \;\mathrm d c - \int\limits_0^1 c D_{ij}^{\mathrm{ens}} \frac{\partial^2 P}{\partial x_i \partial x_j} \;\mathrm d c = \int\limits_0^1 c \frac{\partial}{\partial c}\left[\chi (c - \langle C \rangle) P \right] \mathrm d c \; .
\end{equation}
The order of integration and derivation is swapped on the left hand side and on the right hand side the product rule is applied:

\begin{align}
 \frac{\partial}{\partial t} &\int\limits_0^1 c P \;\mathrm d c + \langle V_i \rangle \frac{\partial}{\partial x_i} \int\limits_0^1 c P \;\mathrm d c - D_{ij}^{\mathrm{ens}} \frac{\partial^2}{\partial x_i \partial x_j} \int\limits_0^1 c P \;\mathrm d c \notag\\
 = &\int\limits_0^1 \left\{ \frac{\partial}{\partial c}\left[ c \chi \left(c - \langle C \rangle \right) P \right] - \chi \left( c - \langle C \rangle \right) P \frac{\partial c}{\partial c} \right\}\;\mathrm d c \; .
\end{align}
On the left hand side, the definition of the mean concentration \(\langle C \rangle := \int c P \;\mathrm d c\) can already be inserted and on the right hand side, the integral is evaluated:

\begin{equation}
 \frac{\partial \langle C \rangle}{\partial t} + \langle V_i \rangle \frac{\partial \langle C \rangle}{\partial x_i} - D_{ij}^{\mathrm{ens}} \frac{\partial^2 \langle C \rangle}{\partial x_i \partial x_j} = \left. c \chi \left(c - \langle C \rangle \right) P \right|_{c=0}^1 - \chi \left(\langle C \rangle - \langle C \rangle \right) P \; .
\end{equation}
Both terms on the right hand side vanish, the second one is obvious, but the first one needs further comment. The case \(c=0\) is clear, but for \(c=1\), the term could potentially result in a non-zero value, if all concentration is gathered at one singular point as a Dirac delta function. But this case is excluded as it is not relevant if studying natural systems. Thus, the transport equation for the mean concentration is

\begin{equation}\label{eq:mean_c_eq_appendix}
 \frac{\partial \langle C \rangle}{\partial t} + \langle V_i \rangle \frac{\partial \langle C \rangle}{\partial x_i} - D_{ij}^{\mathrm{ens}} \frac{\partial^2 \langle C \rangle}{\partial x_i \partial x_j} = 0 \; .
\end{equation}
It can be seen that the mixing term does not influence the mean concentration, as it cancels itself out.

The variance is defined by \eqref{eq:var_def}. Thus, as with the derivation of the mean concentration, we start again from the PDF transport equation \eqref{eq:PDF_transport}, but now we multiply it with \(c^2\) and integrate over the whole concentration space. The order of integration and derivation is swapped and the product rule is applied to the right hand side:

\begin{align}
 \frac{\partial}{\partial t} &\int_0^1 c^2 P \;\mathrm d c + \langle V_i \rangle \frac{\partial}{\partial x_i} \int_0^1 c^2 P \;\mathrm d c - D_{ij}^{\mathrm{ens}} \frac{\partial^2}{\partial x_i \partial x_j} \int_0^1 c^2 P \;\mathrm d c \notag\\
 = &\int_0^1 \left\{ \frac{\partial}{\partial c}\left[ c^2 \chi \left(c - \langle C \rangle \right) P \right] - 2 c \chi \left( c - \langle C \rangle \right) P \right\}\;\mathrm d c \; .
\end{align}
The first term on the right hand side vanishes for the same reason as in the derivation of the mean concentration. The definition of the mean concentration \eqref{eq:mean_c_def} can be inserted into the second term on the right hand side:

\begin{align}
 &\frac{\partial}{\partial t} \int_0^1 c^2 P \;\mathrm d c + \langle V_i \rangle \frac{\partial}{\partial x_i} \int_0^1 c^2 P \;\mathrm d c - D_{ij}^{\mathrm{ens}} \frac{\partial^2}{\partial x_i \partial x_j} \int_0^1 c^2 P \;\mathrm d c \notag\\
 = &-2 \chi \left[ \int_0^1 c^2 P \,\mathrm d c - \langle C \rangle^2 \right] \; .\label{eq:var_der_step}
\end{align}
The term inside the squared brackets on the right hand side could already be replaced by the concentration variance \eqref{eq:var_def}, but to do this for every term, the transport equation of \(\langle C \rangle^2\) needs to be subtracted from equation \eqref{eq:var_der_step}. Therefore, the equation of the squared mean concentration needs to be derived first. This is done by multiplying equation \eqref{eq:mean_c_eq_appendix} by \(\langle C \rangle\), yielding

\begin{equation}\label{eq:c-term-c}
 \langle C \rangle \frac{\partial \langle C \rangle}{\partial t} + \langle C \rangle \langle V_i \rangle \frac{\partial \langle C \rangle}{\partial x_i} - \langle C \rangle D_{ij}^{\mathrm{ens}} \frac{\partial^2 \langle C \rangle}{\partial x_i \partial x_j} = 0 \; .
\end{equation}
By making extensive use of the product rule we arrive at

\begin{align}
 &\frac{\partial \langle C \rangle^2}{\partial t} - \langle C \rangle \frac{\partial \langle C \rangle}{\partial t} + \langle V_i \rangle \frac{\partial \langle C \rangle^2}{\partial x_i} - \langle C \rangle \langle V_i \rangle \frac{\partial \langle C \rangle}{\partial x_i} \notag\\
 - &D_{ij}^{\mathrm{ens}} \left[ \frac{\partial}{\partial x_i} \left(\langle C \rangle \frac{\partial \langle C \rangle}{\partial x_j}\right) - \frac{\partial \langle C \rangle}{\partial x_i} \frac{\partial \langle C \rangle}{\partial x_j} \right] = 0 \; .\label{eq:c_sqrd_step}
\end{align}
The dispersion term is further modified by using the product rule:

\begin{align}
 &D_{ij}^{\mathrm{ens}} \left[ \frac{\partial}{\partial x_i} \left(\langle C \rangle \frac{\partial \langle C \rangle}{\partial x_j}\right) - \frac{\partial \langle C \rangle}{\partial x_i} \frac{\partial \langle C \rangle}{\partial x_j} \right] \notag\\
 =&D_{ij}^{\mathrm{ens}} \left[ \frac{\partial}{\partial x_i} \left( \frac{\partial \langle C \rangle^2}{\partial x_j} - \langle C \rangle \frac{\partial \langle C \rangle}{\partial x_j} \right) - \frac{\partial \langle C \rangle}{\partial x_i} \frac{\partial \langle C \rangle}{\partial x_j} \right] \notag\\
 =&D_{ij}^{\mathrm{ens}} \left[ \frac{\partial^2 \langle C \rangle^2}{\partial x_i \partial x_j} - \langle C \rangle \frac{\partial^2 \langle C \rangle}{\partial x_i \partial x_j} - 2 \frac{\partial \langle C \rangle}{\partial x_i} \frac{\partial \langle C \rangle}{\partial x_j} \right] \; .
\end{align}
Hence, equation \eqref{eq:c_sqrd_step} is transformed into

\begin{align}
 &\frac{\partial \langle C \rangle^2}{\partial t} + \langle V_i \rangle \frac{\partial \langle C \rangle^2}{\partial x_i} - D_{ij}^{\mathrm{ens}} \frac{\partial^2 \langle C \rangle^2}{\partial x_i \partial x_j} + 2 D_{ij}^{\mathrm{ens}} \frac{\partial \langle C \rangle}{\partial x_i} \frac{\partial \langle C \rangle}{\partial x_j} \notag\\
 - &\langle C \rangle \frac{\partial \langle C \rangle}{\partial t} - \langle C \rangle \langle V_i \rangle \frac{\partial \langle C \rangle}{\partial x_i} + \langle C \rangle D_{ij}^{\mathrm{ens}} \frac{\partial^2 \langle C \rangle}{\partial x_i \partial x_j} = 0 \; .\label{eq:c_sqrd_step2}
\end{align}
If we compare the second line of equation \eqref{eq:c_sqrd_step2} with equation \eqref{eq:c-term-c}, we see that it vanishes and the transport equation of \(\langle C \rangle^2\) is

\begin{equation}\label{eq:c_sqrd}
 \frac{\partial \langle C \rangle^2}{\partial t} + \langle V_i \rangle \frac{\partial \langle C \rangle^2}{\partial x_i} - D_{ij}^{\mathrm{ens}} \frac{\partial^2 \langle C \rangle^2}{\partial x_i \partial x_j} + 2 D_{ij}^{\mathrm{ens}} \frac{\partial \langle C \rangle}{\partial x_i} \frac{\partial \langle C \rangle}{\partial x_j} = 0
\end{equation}
Now, equation \eqref{eq:c_sqrd} can be subtracted from equation \eqref{eq:var_der_step}:

\begin{align}
 &\frac{\partial}{\partial t} \left[ \int_0^1 c^2 P \;\mathrm d c - \langle C \rangle^2 \right] + \langle V_i \rangle \frac{\partial}{\partial x_i} \left[ \int_0^1 c^2 P \;\mathrm d c - \langle C \rangle^2 \right] \notag\\
 - &D_{ij}^{\mathrm{ens}} \frac{\partial^2}{\partial x_i \partial x_j} \left[ \int_0^1 c^2 P \;\mathrm d c - \langle C \rangle^2 \right] \notag\\
 = &2 D_{ij}^{\mathrm{ens}} \frac{\partial \langle C \rangle}{\partial x_i} \frac{\partial \langle C \rangle}{\partial x_j} - 2 \chi \left[ \int_0^1 c^2 P \;\mathrm d c - \langle C \rangle^2 \right] \; .
\end{align}
Finally, the definition of the concentration variance \eqref{eq:var_def} is inserted, which yields the transport equation for the variance:

\begin{equation}
 \frac{\partial \sigma_c^2}{\partial t} + \langle V_i \rangle \frac{\partial \sigma_c^2}{\partial x_i} - D_{ij}^{\mathrm{ens}} \frac{\partial^2 \sigma_c^2}{\partial x_i \partial x_j} = 2 D_{ij}^{\mathrm{ens}} \frac{\partial \langle C \rangle}{\partial x_i} \frac{\partial \langle C \rangle}{\partial x_j} - 2 \chi \sigma_c^2 \; .
\end{equation}

\section{Analytical solution of the Variance Transport Equation}\label{app:var-ana-deriv}

An analytical solution of the variance transport equation \eqref{eq:pdf-2-var} will now be derived. As this equation is a linear inhomogeneous partial differential equation, a fundamental solution is derived and convolved with the inhomogeneity of equation \eqref{eq:pdf-2-var} in order to derive the general solution. This way, an analytical solution can be found without making any approximations or assumptions. This derivation is similar to the one presented by Kapoor and Gelhar \cite{kapoor_transport_1994-1}, but we believe the derivation presented here is easier to comprehend and we include a time dependency of the variance decay coefficient \(\chi(t)\).

If we define the differential operator \(L(\mathbf x, t)\) by

\begin{equation}
 L \sigma_c^2 = \frac{\partial \sigma_c^2}{\partial t} + \langle V_i \rangle \frac{\partial \sigma_c^2}{\partial x_i} - D_{ij}^{\mathrm{ens}} \frac{\partial^2 \sigma_c^2}{\partial x_i \partial x_j} - 2 \chi \sigma_c^2 = 0 \; ,
\end{equation}
with the inhomogeneity

\begin{equation}\label{eq:inhom}
 g(\mathbf x, t) = 2 D_{ij}^{\mathrm{ens}} \frac{\partial \langle C \rangle}{\partial x_i} \frac{\partial \langle C \rangle}{\partial x_j} \; ,
\end{equation}
then we can rewrite the concentration variance transport equation \eqref{eq:pdf-2-var} as

\begin{equation}\label{eq:L-sigma}
 L(\mathbf x, t) \sigma_c^2(\mathbf x, t) = g(\mathbf x, t).
\end{equation}
The fundamental solution (or Green's function) \(G(\mathbf x - \mathbf x', t, t')\) is defined as the solution of the differential operator \(L(\mathbf x, t)\) with delta functions as the inhomogeneity:

\begin{equation}\label{eq:fund-sol}
 L(\mathbf x, t) G(\mathbf x - \mathbf x', t, t') = \delta(\mathbf x - \mathbf x') \delta(t - t') \; .
\end{equation}
The fundamental solution is translation invariant in space, because the operator \(L\) has constant coefficients with respect to \(\mathbf x\).
The general solution of equation \eqref{eq:pdf-2-var} is given by

\begin{equation}\label{eq:var-green}
 \sigma_c^2(\mathbf x, t) = \sigma_{c_h}^2(\mathbf x, t) + \int_0^t \int_{\mathbb{R}^d} G(\mathbf x - \mathbf x', t, t') g(\mathbf x', t') \mathrm d \mathbf x' \mathrm d t' \; ,
\end{equation}
where \(\sigma_{c_h}^2(\mathbf x, t)\) is the solution of equation \eqref{eq:pdf-2-var} without the inhomogeneity. But because we assume that the initial condition is known exactly, the variance at time \(t = 0\) is \(\sigma_c^2(\mathbf x, t=0) = 0\). Thus, without the inhomogeneity, which acts as the only source term, the solution of the homogeneous partial differential equation is \( \sigma_{c_h}^2(\mathbf x, t) = 0 \) for all times. Therefore, the solution of the homogeneous equation can be dropped.

If Green's function is known, the solution of equation \eqref{eq:pdf-2-var} can be calculated from equation \eqref{eq:var-green}, which is a convolution of Green's function and the inhomogeneity in physical space:

\begin{align}\label{eq:var-green2}
 \sigma_c^2(\mathbf x, t) = &\int_0^t \int_{\mathbb{R}^d} G(\mathbf x - \mathbf x', t, t') g(\mathbf x', t') \; \mathrm d \mathbf x' \mathrm d t' \notag\\
 = &\int_0^t (G \ast g)(\mathbf x, t, t') \; \mathrm d t' \notag\\
 = &\int_0^t \mathcal{F}^{-1} \left[ \tilde G(\mathbf k, t, t') \tilde g(\mathbf k, t') \right] \mathrm d t' \; ,
\end{align}
where \(\mathcal{F}^{-1}\) denotes the inverse Fourier transform and a tilde denotes the Fourier transform of a function. Hence, \(\tilde G\) and \(\tilde g\) need to be calculated in order to obtain the solution \(\sigma_c^2\).

Fourier transforming both sides of equation \eqref{eq:fund-sol} gives an inhomogeneous ordinary differential equation in the frequency domain:

\begin{equation}
 \left( \frac{\partial}{\partial t} + I \langle V_i \rangle k_i + D_{ij}^{\mathrm{ens}} k_i k_j + 2 \chi(t) \right) \tilde G(\mathbf k, t, t') = \delta(t-t') \; ,
\end{equation}
with \(I\) being the imaginary unit. This ordinary differential equation can be solved by separation of variables for a solution of the homogeneous equation. But with the initial condition of \(\sigma_c^2(\mathbf x, t=0) = 0\) as explained above and thus \(G(\mathbf x - \mathbf x', t=0, t') = 0\), the homogeneous solution always stays zero, as the inhomogeneity is the only source term. Nevertheless, the homogeneous solution can be used as a starting point to guess a particular solution of the inhomogeneous solution, yielding

\begin{equation}\label{eq:G-tilde}
 \tilde G(\mathbf k, t, t') = \Theta(t-t') \exp\left(-\left( D_{ij}^{\mathrm{ens}} k_i k_j + I \langle V_i \rangle k_i \right)(t-t') -2 \int_{t'}^t \mathrm d t'' \chi(t'') \right) ,
\end{equation}
where \(\Theta\) is the Heaviside step function. In order to transform the inhomogeneity \eqref{eq:inhom}, the transformed mean concentration \(\langle C \rangle\) from equation \eqref{eq:mean-c-ana} needs to be plugged in:

\begin{equation}
 \tilde g(\mathbf k, t) = \mathcal{F}\left[ 2 D_{ij}^{\mathrm{ens}} \frac{\partial \langle C \rangle}{\partial x_i} \frac{\partial \langle C \rangle}{\partial x_j} \right] = \frac{-2 D_{ij}^{\mathrm{ens}}}{(2\pi)^{d/2}} \; k_i \tilde{\langle C \rangle} \ast k_j \tilde{\langle C \rangle} \; .
\end{equation}
At this point, the time shift \(t_0\) is needed. Otherwise, a singularity for \(t=0\) would cause problems, as the Gaussian distribution would tend to a Dirac delta function for small times. The Fourier transformed mean concentration is

\begin{equation}
 \tilde{\langle C \rangle} = \frac{1}{(2\pi)^{d/2}} \exp\left( -D_{ij}^{\mathrm{ens}} k_i k_j (t+t_0) - I k_i \langle V_i \rangle t \right) \; .
\end{equation}
With this solution, the Fourier transformed inhomogeneity can be calculated:

\begin{align}
 \tilde g(\mathbf k, t) = &\frac{D_{ij}^{\mathrm{ens}}}{2 (2\pi)^d} \frac{1}{(2 D_{ij}^{\mathrm{ens}} (t+t_0))^{d/2}} \left[ \frac{d}{D_{ij}^{\mathrm{ens}} (t+t_0)} - k_i k_j \right] \notag\\
 &\exp\left( -\frac{1}{2} D_{ij}^{\mathrm{ens}} (t+t_0) k_i k_j - I \langle V_i \rangle k_i t \right) \; .\label{eq:g-tilde}
\end{align}
Finally, the transformed Green's function \eqref{eq:G-tilde} and the transformed inhomogeneity \eqref{eq:g-tilde} are inserted into equation \eqref{eq:var-green2}:

\begin{align}
 \sigma_c^2 &(\mathbf x, t) = \frac{D_{ij}^{\mathrm{ens}}}{2 (2 \pi)^{3d/2}} \int_0^t \mathrm d t' \frac{\Theta(t-t')}{[2 D_{ij}^{\mathrm{ens}} (t' + t_0)]^{d/2}} \int_{\mathbb{R}^d} \mathrm d \mathbf k \left[ \frac{d}{D_{ij}^{\mathrm{ens}} (t' + t_0)} - k_i k_j \right] \notag\\
 &\exp\left( -\frac{1}{2} D_{ij}^{\mathrm{ens}} \left(2t - t' + t_0 \right) k_i k_j + I (x_i - \langle V_i \rangle t) k_i - 2 \int_{t'}^t \mathrm d t'' \chi(t'') \right).
\end{align}
By completing the square for the variable \(\mathbf k\), the Fourier integrand is lead back to a Gaussian function which can be integrated. Because the ensemble dispersion tensor is diagonal, we can simplify the expression by only considering the diagonal elements. Now, only a final time integral remains to be calculated:

\begin{align}
 \sigma_c^2 &(\mathbf x, t) = \sum_{i=1}^d 2 D_{ii}^{\mathrm{ens}} \int_0^t \mathrm d t' \prod_{j=1}^d \frac{\exp\left( -\frac{(x_j - \langle V_j \rangle t)^2}{2 D_{jj}^{\mathrm{ens}} (2t + t_0 - t')} \right)} {\left[(2 \pi D_{jj}^{\mathrm{ens}})^2 (2t + t_0 - t') (t'+t_0) \right]^{1/2}} \notag\\
 &\left[ \frac{t-t'}{2 D_{ii}^{\mathrm{ens}} (2t + t_0 - t')(t'+t_0)} + \frac{(x_i - \langle V_i \rangle t)^2}{\left(2 D_{ii}^{\mathrm{ens}} (2t + t_0 - t')\right)^2} \right] \notag\\
 &\exp\left( -2 \int_{t'}^t \mathrm d t'' \chi(t'') \right) \; .
\end{align}
This integral can either be evaluated analytically by using a long-time approximation or by applying numerical methods.

\bibliographystyle{elsarticle-num}
\bibliography{paper}

\begin{thebibliography}{10}
\expandafter\ifx\csname url\endcsname\relax
  \def\url#1{\texttt{#1}}\fi
\expandafter\ifx\csname urlprefix\endcsname\relax\def\urlprefix{URL }\fi
\expandafter\ifx\csname href\endcsname\relax
  \def\href#1#2{#2} \def\path#1{#1}\fi

\bibitem{WWAP_united_2012}
{WWAP}, The {United} {Nations} {World} {Water} {Development} {Report} 4:
  {Managing} {Water} under {Uncertainty} and {Risk}, no. Vol. 1 in World
  {Water} {Assessment} {Programme}, Unesco, Paris, 2012.

\bibitem{gelhar_three_1983}
L.~W. Gelhar, C.~L. Axness, Three {Dimensional} {Stochastic} {Analysis} of
  {Macrodispersion} in {Aquifers}, Water Resour. Res. 19~(1) (1983) 161--180.
\newblock \href {http://dx.doi.org/10.1029/WR019i001p00161}
  {\path{doi:10.1029/WR019i001p00161}}.

\bibitem{burr_nonreactive_1994}
D.~T. Burr, E.~A. Sudicky, R.~L. Naff, Nonreactive and reactive solute
  transport in three-dimensional heterogeneous porous media: {Mean}
  displacement, plume spreading, and uncertainty, Water Resour. Res. 30~(3)
  (1994) 791--815.
\newblock \href {http://dx.doi.org/10.1029/93WR02946}
  {\path{doi:10.1029/93WR02946}}.

\bibitem{tennekes_first_1972}
H.~Tennekes, J.~L. Lumley, A {First} {Course} in {Turbulence}, MIT Press,
  Cambridge, Mass., 1972.

\bibitem{dentz_temporal_2000-1}
M.~Dentz, H.~Kinzelbach, S.~Attinger, W.~Kinzelbach, Temporal behavior of a
  solute cloud in a heterogeneous porous medium 1 .{Point}-like injection,
  Water Resour. Res. 36~(12) (2000) 3591--3604.
\newblock \href {http://dx.doi.org/10.1029/2000WR900162}
  {\path{doi:10.1029/2000WR900162}}.

\bibitem{andricevic_effects_1998}
R.~Andričević, Effects of local dispersion and sampling volume on the
  evolution of concentration fluctuations in aquifers, Water Resour. Res.
  34~(5) (1998) 1115--1129.
\newblock \href {http://dx.doi.org/10.1029/98WR00260}
  {\path{doi:10.1029/98WR00260}}.

\bibitem{kapoor_transport_1994}
V.~Kapoor, L.~W. Gelhar, Transport in three-dimensionally heterogeneous
  aquifers: 1. {Dynamics} of concentration fluctuations, Water Resour. Res.
  30~(6) (1994) 1775--1788.
\newblock \href {http://dx.doi.org/10.1029/94WR00076}
  {\path{doi:10.1029/94WR00076}}.

\bibitem{kapoor_transport_1994-1}
V.~Kapoor, L.~W. Gelhar, Transport in three-dimensionally heterogeneous
  aquifers 2. {Predictions} and observations of concentration fluctuations,
  Water Resour. Res. 30~(6) (1994) 1789--1801.
\newblock \href {http://dx.doi.org/10.1029/94WR00075}
  {\path{doi:10.1029/94WR00075}}.

\bibitem{dagan_stochastic_1982}
G.~Dagan, Stochastic {Modeling} of {Groundwater} {Flow} by {Unconditional} and
  {Conditional} {Probabilities} 1.{Conditional} {Simulation} and the {Direct}
  {Problem}, Water Resour. Res. 18~(4) (1982) 813--833.
\newblock \href {http://dx.doi.org/10.1029/WR018i004p00835}
  {\path{doi:10.1029/WR018i004p00835}}.

\bibitem{kapoor_advection-diffusion_1997}
V.~Kapoor, P.~K. Kitanidis, Advection-diffusion in spatially random flows:
  {Formulation} of concentration covariance, Stoch. Hydrol. Hydraul. 11~(5)
  (1997) 397--422.
\newblock \href {http://dx.doi.org/10.1007/BF02427926}
  {\path{doi:10.1007/BF02427926}}.

\bibitem{andricevic_evaluation_1996}
R.~Andričević, V.~Cvetković, Evaluation of {Risk} from {Contaminants}
  {Migrating} by {Groundwater}, Water Resour. Res. 32~(3) (1996) 611--621.
\newblock \href {http://dx.doi.org/10.1029/95WR03530}
  {\path{doi:10.1029/95WR03530}}.

\bibitem{de_barros_simple_2011}
F.~P.~J. de~Barros, A.~Fiori, A.~Bellin, A simple closed-form solution for
  assessing concentration uncertainty, Water Resour. Res. 47~(12) (2011) 1--5.
\newblock \href {http://dx.doi.org/10.1029/2011WR011107}
  {\path{doi:10.1029/2011WR011107}}.

\bibitem{fiorotto_solute_2002}
V.~Fiorotto, E.~Caroni, Solute concentration statistics in heterogeneous
  aquifers for finite {Peclet} values, Transp. Porous Media 48~(3) (2002)
  331--351.
\newblock \href {http://dx.doi.org/10.1023/A:1015744421033}
  {\path{doi:10.1023/A:1015744421033}}.

\bibitem{srzic_impact_2013}
V.~Srzic, V.~Cvetkovic, R.~Andricevic, H.~Gotovac, Impact of aquifer
  heterogeneity structure and local-scale dispersion on solute concentration
  uncertainty: {Impact} of {Aquifer} {Heterogeneity} on {Concentration}
  {Uncertainty}, Water Resour. Res. 49~(6) (2013) 3712--3728.
\newblock \href {http://dx.doi.org/10.1002/wrcr.20314}
  {\path{doi:10.1002/wrcr.20314}}.

\bibitem{celis_lagrangian_2015}
C.~Celis, L.~F. Figueira~da Silva, Lagrangian {Mixing} {Models} for {Turbulent}
  {Combustion}: {Review} and {Prospects}, Flow, Turbul. Combust. 94~(3) (2015)
  643--689.
\newblock \href {http://dx.doi.org/10.1007/s10494-015-9597-1}
  {\path{doi:10.1007/s10494-015-9597-1}}.

\bibitem{meyer_joint_2010}
D.~W. Meyer, P.~Jenny, H.~A. Tchelepi, A joint velocity-concentration {PDF}
  method for tracer flow in heterogeneous porous media, Water Resour. Res.
  46~(12) (2010) 1--17.
\newblock \href {http://dx.doi.org/10.1029/2010WR009450}
  {\path{doi:10.1029/2010WR009450}}.

\bibitem{pope_pdf_1985}
S.~B. Pope, {PDF} {Methods} for {Turbulent} {Reactive} {Flows}, Prog. Energy
  Combust. Sci. 11~(2) (1985) 119--192.
\newblock \href {http://dx.doi.org/10.1016/0360-1285(85)90002-4}
  {\path{doi:10.1016/0360-1285(85)90002-4}}.

\bibitem{fox_computational_2003}
R.~O. Fox, Computational {Models} for {Turbulent} {Reacting} {Flows}, Cambridge
  {Series} in {Chemical} {Engineering}, Cambridge University Press, New York,
  2003.

\bibitem{suciu_fokker-planck_2015}
N.~Suciu, F.~A. Radu, S.~Attinger, L.~Schüler, P.~Knabner, A {Fokker}-{Planck}
  approach for probability distributions of species concentrations transported
  in heterogeneous media, J. Comput. Appl. Math. 289 (2015) 241--252.
\newblock \href {http://dx.doi.org/10.1016/j.cam.2015.01.030}
  {\path{doi:10.1016/j.cam.2015.01.030}}.

\bibitem{suciu_consistency_2015}
N.~Suciu, L.~Schüler, S.~Attinger, C.~Vamos, P.~Knabner, Consistency issues in
  {PDF} methods, An. St. Univ. Ovidius Constanta, Ser. Mat. 23~(3) (2015)
  187--208.
\newblock \href {http://dx.doi.org/10.1515/auom-2015-0055}
  {\path{doi:10.1515/auom-2015-0055}}.

\bibitem{suciu_towards_????}
N.~Suciu, L.~Schüler, S.~Attinger, P.~Knabner, Towards a filtered density
  function approach for reactive transport in groundwater, Adv. Water
  Resour.Accepted.

\bibitem{villermaux_representation_1972}
J.~Villermaux, J.~C. Devillon, Représentation de la coalescence et de la
  redispersion des domaines de ségrégation dans un fluide par un modèle
  d’interaction phénoménologique., in: Proceedings of the 2nd
  {International} symposium on chemical reaction engineering, Elsevier New
  York, 1972, pp. 1--13.

\bibitem{dopazo_approach_1974}
C.~Dopazo, E.~E. O'Brien, An approach to autoignition of a turbulent mixture,
  Acta Astronaut. 1 (1974) 1239--1266.
\newblock \href {http://dx.doi.org/10.1016/0094-5765(74)90050-2}
  {\path{doi:10.1016/0094-5765(74)90050-2}}.

\bibitem{colucci_filtered_1998}
P.~J. Colucci, F.~A. Jaberi, P.~Givi, S.~B. Pope, Filtered density function for
  large eddy simulation of turbulent reacting flows, Phys. Fluids 10~(2) (1998)
  499--515.
\newblock \href {http://dx.doi.org/10.1063/1.869537}
  {\path{doi:10.1063/1.869537}}.

\bibitem{raman_consistent_2007}
V.~Raman, H.~Pitsch, A consistent {LES}/filtered-density function formulation
  for the simulation of turbulent flames with detailed chemistry, Proc.
  Combust. Inst. 31~(2) (2007) 1711--1719.
\newblock \href {http://dx.doi.org/10.1016/j.proci.2006.07.152}
  {\path{doi:10.1016/j.proci.2006.07.152}}.

\bibitem{popov_implicit_2014}
P.~P. Popov, S.~B. Pope, Implicit and explicit schemes for mass consistency
  preservation in hybrid particle/finite-volume algorithms for turbulent
  reactive flows, J. Comput. Phys. 257 (2014) 352--373.
\newblock \href {http://dx.doi.org/10.1016/j.jcp.2013.09.005}
  {\path{doi:10.1016/j.jcp.2013.09.005}}.

\bibitem{sabelnikov_extended_2006}
V.~Sabel'nikov, M.~Gorokhovski, N.~Baricault, The extended {IEM} mixing model
  in the framework of the composition {PDF} approach: applications to diesel
  spray combustion, Combust. Theory Modell. 10~(1) (2006) 155--169.
\newblock \href {http://dx.doi.org/10.1080/13647830500348109}
  {\path{doi:10.1080/13647830500348109}}.

\bibitem{jones_les_2012}
W.~Jones, A.~Marquis, V.~Prasad, {LES} of a turbulent premixed swirl burner
  using the {Eulerian} stochastic field method, Combust. Flame 159~(10) (2012)
  3079--3095.
\newblock \href {http://dx.doi.org/10.1016/j.combustflame.2012.04.008}
  {\path{doi:10.1016/j.combustflame.2012.04.008}}.

\bibitem{dodoulas_large_2013}
I.~A. Dodoulas, S.~Navarro-Martinez, Large {Eddy} {Simulation} of {Premixed}
  {Turbulent} {Flames} {Using} the {Probability} {Density} {Function}
  {Approach}, Flow, Turbul. Combust. 90~(3) (2013) 645--678.
\newblock \href {http://dx.doi.org/10.1007/s10494-013-9446-z}
  {\path{doi:10.1007/s10494-013-9446-z}}.

\bibitem{suciu_diffusion_2014}
N.~Suciu, Diffusion in random velocity fields with applications to contaminant
  transport in groundwater, Adv. Water Resour. 69 (2014) 114--133.
\newblock \href {http://dx.doi.org/10.1016/j.advwatres.2014.04.002}
  {\path{doi:10.1016/j.advwatres.2014.04.002}}.

\bibitem{dentz_temporal_2002}
M.~Dentz, H.~Kinzelbach, S.~Attinger, W.~Kinzelbach, Temporal behavior of a
  solute cloud in a heterogeneous porous medium 3. {Numerical} simulations,
  Water Resour. Res. 38~(7) (2002) 23--1--23--13.
\newblock \href {http://dx.doi.org/10.1029/2001WR000436}
  {\path{doi:10.1029/2001WR000436}}.

\bibitem{vamos_generalized_2003}
C.~Vamoş, N.~Suciu, H.~Vereecken, Generalized random walk algorithm for the
  numerical modeling of complex diffusion processes, J. Comput. Phys. 186~(2)
  (2003) 527--544.
\newblock \href {http://dx.doi.org/10.1016/S0021-9991(03)00073-1}
  {\path{doi:10.1016/S0021-9991(03)00073-1}}.

\bibitem{kraichnan_diffusion_1970}
R.~H. Kraichnan, Diffusion by a {Random} {Velocity} {Field}, Phys. Fluids
  13~(1) (1970) 22--31.
\newblock \href {http://dx.doi.org/10.1063/1.1692799}
  {\path{doi:10.1063/1.1692799}}.

\bibitem{hese_generating_2014}
F.~Heße, V.~Prykhod'ko, S.~Schlüter, S.~Attinger, Generating random fields
  with a truncated power-law variogram. {A} comparison of several numerical
  methods with respect to accurary and reproduction of structural features.,
  Environ. Model. Softw. 55 (2014) 32--48.
\newblock \href {http://dx.doi.org/10.1016/j.envsoft.2014.01.013}
  {\path{doi:10.1016/j.envsoft.2014.01.013}}.

\bibitem{eberhard_self-averaging_2007}
J.~P. Eberhard, N.~Suciu, C.~Vamoş,
  \href{http://iopscience.iop.org/1751-8121/40/4/002}{On the self-averaging of
  dispersion for transport in quasi-periodic random media}, J. Phys. A: Math.
  Gen. 40~(4) (2007) 597.
\newblock \href {http://dx.doi.org/10.1088/1751-8113/40/4/002}
  {\path{doi:10.1088/1751-8113/40/4/002}}.
\newline\urlprefix\url{http://iopscience.iop.org/1751-8121/40/4/002}

\bibitem{drummond_scalar_1984}
I.~T. Drummond, S.~Duane, R.~R. Horgan, Scalar diffusion in simulated helical
  turbulence with molecular diffusivity, J. Fluid Mech. 138 (1984) 75--91.
\newblock \href {http://dx.doi.org/10.1017/S0022112084000045}
  {\path{doi:10.1017/S0022112084000045}}.

\bibitem{suciu_numerical_2006}
N.~Suciu, C.~Vamoş, J.~Vanderborght, H.~Hardelauf, H.~Vereecken, Numerical
  {Investigations} on {Ergodicity} of {Solute} {Transport} in {Heterogeneous}
  {Aquifers}, Water Resour. Res. 42~(4) (2006) 1--17.
\newblock \href {http://dx.doi.org/10.1029/2005WR004546}
  {\path{doi:10.1029/2005WR004546}}.

\bibitem{suciu_coupled_2013}
N.~Suciu, F.~A. Radu, A.~Prechtel, F.~Brunner, P.~Knabner, A coupled finite
  element–global random walk approach to advection-dominated transport in
  porous media with random hydraulic conductivity, J. Comput. Appl. Math. 246
  (2013) 27--37.
\newblock \href {http://dx.doi.org/10.1016/j.cam.2012.06.027}
  {\path{doi:10.1016/j.cam.2012.06.027}}.

\bibitem{im_large_1997}
H.~G. Im, T.~S. Lund, J.~H. Ferziger,
  \href{http://scitation.aip.org/content/aip/journal/pof2/9/12/10.1063/1.869517}{Large
  eddy simulation of turbulent front propagation with dynamic subgrid models},
  Phys. Fluids 9~(12) (1997) 3826--3833.
\newblock \href {http://dx.doi.org/10.1063/1.869517}
  {\path{doi:10.1063/1.869517}}.
\newline\urlprefix\url{http://scitation.aip.org/content/aip/journal/pof2/9/12/10.1063/1.869517}

\bibitem{pierce_dynamic_1998}
C.~D. Pierce, P.~Moin,
  \href{http://scitation.aip.org/content/aip/journal/pof2/10/12/10.1063/1.869832}{A
  dynamic model for subgrid-scale variance and dissipation rate of a conserved
  scalar}, Phys. Fluids 10~(12) (1998) 3041--3044.
\newblock \href {http://dx.doi.org/10.1063/1.869832}
  {\path{doi:10.1063/1.869832}}.
\newline\urlprefix\url{http://scitation.aip.org/content/aip/journal/pof2/10/12/10.1063/1.869832}

\end{thebibliography}
\end{document}